%% file: main.tex
\newif\ifacmversion
\acmversionfalse

\ifacmversion

\documentclass[sigconf]{acmart}

\input{packages}

\input{commands}

\AtBeginDocument{%
  }
    
\copyrightyear{2025}
\acmYear{2025}
\setcopyright{cc}
\setcctype{by}
\acmConference[MICRO '25]{58th IEEE/ACM International Symposium on Microarchitecture}{October 18--22, 2025}{Seoul, Republic of Korea}
\acmBooktitle{58th IEEE/ACM International Symposium on Microarchitecture (MICRO '25), October 18--22, 2025, Seoul, Republic of Korea}
\acmDOI{10.1145/3725843.3756029}
\acmISBN{979-8-4007-1573-0/2025/10}

\setcitestyle{compress}

\else

\documentclass[camera,letterpaper,nomarginnotes,nonarrowgutter]{jpaper}

\input{packages}
\input{commands}
\AtBeginDocument{%
  }

\fancypagestyle{iterationsfirstpage}
{
    \fancyhead{}
    \fancyhead[C]{
    \textcolor{red}{CONFIDENTIAL DRAFT -- DO NOT DISTRIBUTE -- TO APPEAR IN MICRO'25} \\ 
    \emph{\textcolor{blue}{Version \versionnum~---~\today, \ampmtime}}
    }
}

\fi

\newcommand{\affilETH}{$^{1}$}
\newcommand{\affilOguzhan}{$^{1,2}$}
\newcommand{\affilGiray}{$^{1,3}$}

\begin{document}

    \title{Understanding and Mitigating Covert Channel and Side Channel Vulnerabilities Introduced by RowHammer Defenses}

\sloppy

\ifacmversion
\author{F. Nisa Bostanc{\i}}
\affiliation{%
  \institution{ETH Z{\"u}rich}
  \city{Z{\"u}rich}
  \country{Switzerland}
}

\author{Oğuzhan Canpolat}
\affiliation{%
  \institution{ETH Z{\"u}rich}
  \city{Z{\"u}rich}
  \country{Switzerland}
}
\affiliation{%
  \institution{TOBB ET{\"U}}
  \city{Ankara}
  \country{T{\"u}rkiye}
}

\author{Ataberk Olgun}
\affiliation{%
  \institution{ETH Z{\"u}rich}
  \city{Z{\"u}rich}
  \country{Switzerland}
}

\author{{\.I}smail Emir Y{\"u}ksel}
\affiliation{%
  \institution{ETH Z{\"u}rich}
  \city{Z{\"u}rich}
  \country{Switzerland}
}

\author{Konstantinos Kanellopoulos}
\affiliation{%
  \institution{ETH Z{\"u}rich}
  \city{Z{\"u}rich}
  \country{Switzerland}
}

\author{Mohammad Sadrosadati}
\affiliation{%
  \institution{ETH Z{\"u}rich}
  \city{Z{\"u}rich}
  \country{Switzerland}
}

\author{A.~Giray~Ya\u{g}l{\i}k\c{c}{\i}}
\affiliation{%
  \institution{ETH Z{\"u}rich}
  \city{Z{\"u}rich}
  \country{Switzerland}
}
\affiliation{%
  \institution{CISPA}
  \city{Saarbrücken}
  \country{Germany}
}
\author{Onur Mutlu}
\affiliation{%
  \institution{ETH Z{\"u}rich}
  \city{Z{\"u}rich}
  \country{Switzerland}
}

\renewcommand{\shortauthors}{Bostanci et al.}
\else
\author
{{F. Nisa Bostanc\i}\affilETH\qquad%
{Oğuzhan Canpolat}\affilOguzhan\qquad
{Ataberk Olgun}\affilETH\qquad
{İsmail Emir Y\"{u}ksel}\affilETH\qquad\\
{Konstantinos Kanellopoulos}\affilETH\qquad
{Mohammad Sadrosadati}\affilETH\qquad
{A. Giray Ya\u{g}l{\i}k\c{c}{\i}}\affilGiray\qquad%
{Onur Mutlu}\affilETH\qquad\\\vspace{1mm}
{$^{1}${ETH Z{\"u}rich}\quad$^{2}${TOBB ET\"U}\quad$^{3}${CISPA}}}
\fi

%Enables the camera ready header and footer
\ifcamerareadyiterations 
    \ifacmversion
    \thispagestyle{plain} 
    \else
      \thispagestyle{iterationsfirstpage}  
    \fi
    \pagestyle{plain}
    \pagenumbering{arabic}

\else
    \renewcommand{\headrulewidth}{0pt}
    \fancypagestyle{firstpage}{
        \fancyhead{} % clear all header and footer fields
    \renewcommand{\footrulewidth}{0pt}
    }
\thispagestyle{firstpage}
\fi

\setlength{\footskip}{\paperheight
  -(1in+\voffset+\topmargin+\headheight+\headsep+\textheight)
  -0.7in}
\pagenumbering{arabic}
\pagestyle{plain}

\fancypagestyle{firstpage}
{
    \fancyhead{}
    \begin{tikzpicture}[remember picture,overlay]
    \node [xshift=150mm,yshift=-12.5mm]
    at (current page.north west) {\href{https://www.acm.org/publications/policies/artifact-review-and-badging-current}{\includegraphics[width=1.7cm]{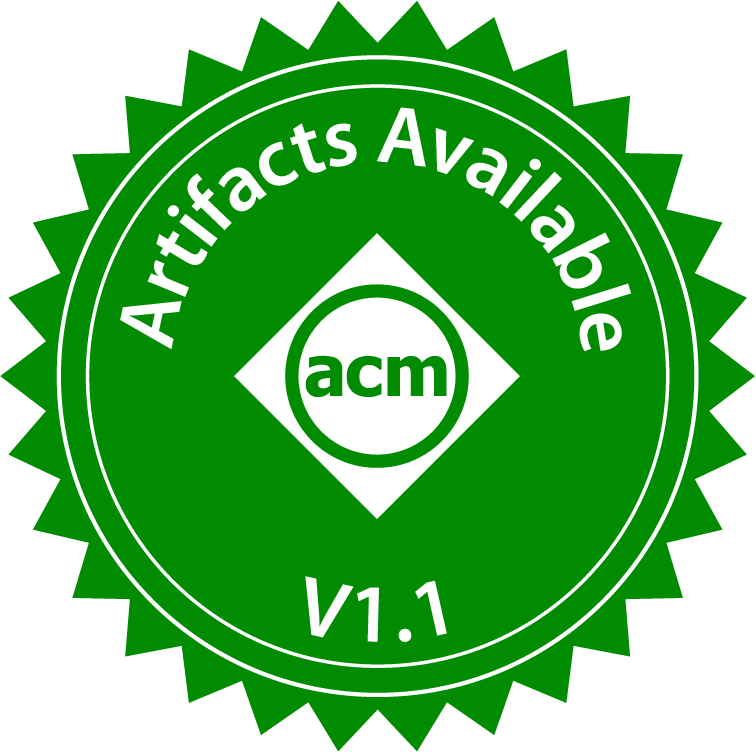}}} ;
    \node [xshift=168mm,yshift=-12.5mm]
    at (current page.north west) {\href{https://www.acm.org/publications/policies/artifact-review-and-badging-current}{\includegraphics[width=1.7cm]{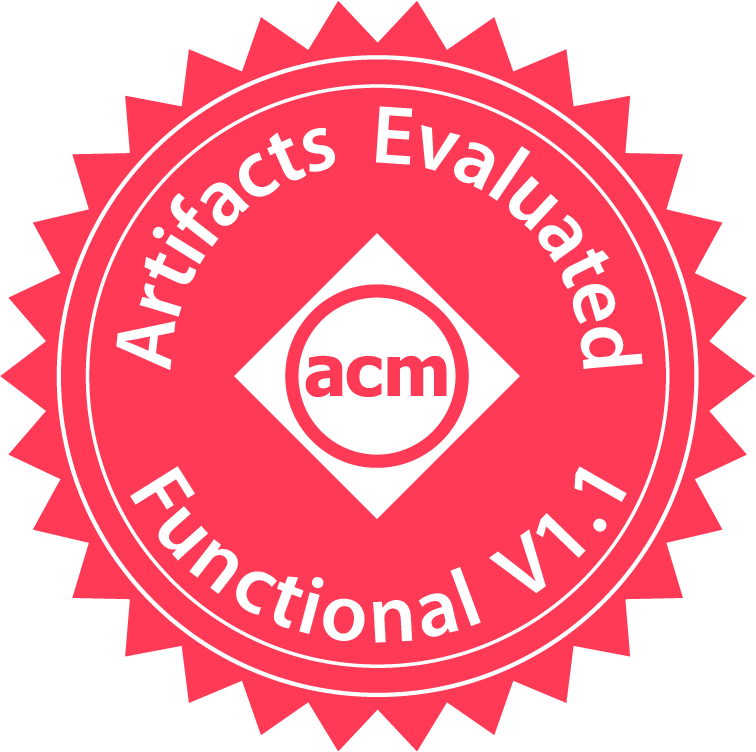}}} ;
    \node [xshift=186mm,yshift=-12.5mm]
    at (current page.north west) {\href{https://www.acm.org/publications/policies/artifact-review-and-badging-current}{\includegraphics[width=1.7cm]{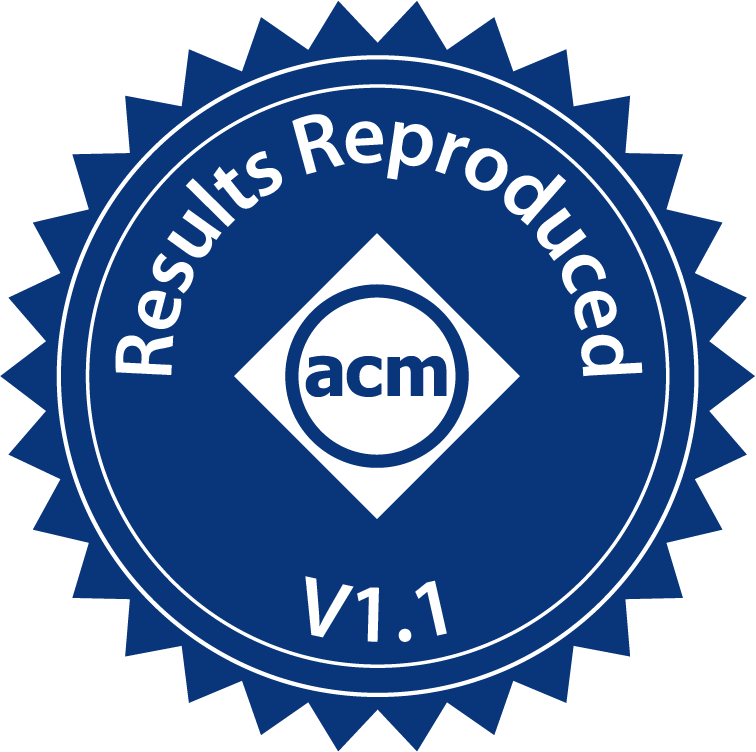}}} ;
    \end{tikzpicture}
  \renewcommand{\headrulewidth}{0pt}
  \pagenumbering{arabic}
  \fancyfoot[C]{\large\thepage}
}

\pagenumbering{arabic}

\newcounter{version}
\setcounter{version}{3}

\ifacmversion
    \begin{abstract}
    \input{sections/00_abstract}
    \end{abstract}
    \maketitle
\else
    \setlength{\droptitle}{-9mm} 
    \maketitle
    \thispagestyle{firstpage}
    \setstretch{0.98}
    \begin{abstract}
    \input{sections/00_abstract}
    \end{abstract}
\fi

\glsresetall
\input{sections/01_intro}

\glsresetall
\setstretch{0.99}
\input{sections/02_background}
\input{sections/03_rhsc}

\input{sections/04_methodology}

\input{sections/05_0_casestudy_prac}

\input{sections/05_1_casestudy}
\input{sections/05_2_casestudy}
\input{sections/12_comparison_new}

\input{sections/10_sensitivitystudy}

\setstretch{0.99}
\input{sections/06_discussion}

\input{sections/07_othermechs}
\setstretch{1}
\input{sections/08_relatedwork}

\input{sections/09_conclusion}

\ifacmversion
\begin{acks}
\else
\section*{Acknowledgments} 
\fi
{An earlier version of this paper was submitted to ISCA 2025 on November 22, 2024, and ultimately, was rejected with  "doubts about [the] performance of this solution (randomly initialized activation counters, \secref{sec:riac})". \nbcr{3}{These "doubts", \nbcr{3}{which were easily addressable, were \emph{not} raised}} in any of the ISCA reviews; therefore, they could not have been rebutted and addressed during the rebuttal period. 
\nbcr{2}{These concerns do not have a significant impact on the results shown in our ISCA 2025 submission, and are nevertheless addressed in this version.}
We thank the anonymous reviewers of {MICRO 2025} for feedback. {We thank the} SAFARI Research Group members (especially Andreas Kosmas Kakolyris, Harsh Songara, and Konstantinos Sgouras) for
{constructive} feedback and the stimulating intellectual {environment.} We
acknowledge the generous gift funding provided by our industrial partners
({especially} Google, Huawei, Intel, Microsoft), which has been instrumental in
enabling the research we have been conducting on read disturbance in DRAM {in
particular and memory systems in
general~\cite{mutlu2023retrospective}.} This work was in part
supported by the Google Security and Privacy Research Award and the Microsoft
Swiss Joint Research Center.}
\ifacmversion
\end{acks}
\fi

\ifacmversion
\else
\bibliographystyle{unsrt}
\fi
\bibliography{refs}

\nobalance
\clearpage
\appendix

\input{sections/20_artifact}
\end{document}

%% file: packages.tex
\ifacmversion
\else
    \usepackage[bookmarks=true,breaklinks=true,colorlinks,linkcolor=black,citecolor=NavyBlue,urlcolor=blue]{hyperref}
\fi
\usepackage{algorithmic}
\usepackage{graphicx}
\usepackage{textcomp}
\usepackage{glossaries} %
\setkeys{glslink}{hyper=false} %
\usepackage{duckuments} %
\usepackage{url}
\usepackage{balance}
\usepackage{siunitx}
\usepackage{multirow}

\usepackage{setspace}
\usepackage{listings}
\lstdefinestyle{customc}{
  breaklines=true,
  frame=single,
  framexleftmargin=10pt, %
  xleftmargin=15pt, %
  xrightmargin=1pt, %
  language=C++,
  morekeywords={uint64_t}, %
  showstringspaces=false,
  basicstyle=\footnotesize\ttfamily,
  keywordstyle=\bfseries\color{black!80!black},
  commentstyle=\itshape\color{orange},
  identifierstyle=\color{black},
  stringstyle=\color{orange},
  numbers=left,
  numberstyle=\tiny\color{gray}, %
  stepnumber=1,
  numbersep=5pt, %
  escapeinside={*@}{@*}, %
  backgroundcolor=\color{white} %
}
\usepackage{fancyhdr}
\usepackage{datetime}
\usepackage{datetime2}

\usepackage{tabularray}
\usepackage{hhline}

\ifacmversion
\else
    \usepackage[sort,compress]{cite}
\fi

\usepackage{mathptmx} % This is Times font
\usepackage{subcaption}

%% file: commands.tex
\newcommand{\X}[0]{{LeakyHammer}}

\newcommand{\rh}[0]{RowHammer}
\input{ref_macros}
\input{glossary}

\newcommand{\secref}[1]{§\ref{#1}}
\newcommand{\figref}[1]{Fig.~\ref{#1}}

\usepackage{tikz}

\newcommand{\head}[1]{{\noindent\textbf{#1.} }}

\definecolor{nbs}{rgb}{0.88, 0.07, 0.37}
\definecolor{iy}{rgb}{0.0, 0.36, 0.05}
\definecolor{gfored}{rgb}{0.580, 0.050, 0.211}
\definecolor{bluepigment}{rgb}{0.2, 0.2, 0.6}
\definecolor{brandeisblue}{rgb}{0.0, 0.44, 1.0}
\definecolor{ouscolor}{rgb}{0.0, 0.2, 0.4}
\definecolor{omc}{rgb}{1, 0.01, 0.01}

\definecolor{ra}{rgb}{0.05, 0.5, 0.06}
\definecolor{rb}{rgb}{0.9, 0.17, 0.31}
\definecolor{rc}{rgb}{0.06, 0.2, 0.65}
\definecolor{rd}{rgb}{0.93, 0.53, 0.38}
\definecolor{re}{rgb}{0.1, 0.5, 0.5}

\definecolor{cq}{rgb}{0.0, 0.5, 1.0}

\newif\ifcamerareadyiterations
\camerareadyiterationsfalse

\newif\ifcameraready
\camerareadytrue

\ifcameraready
  \newcommand\nbcr[2]{#2}
  \newcommand\atbcr[2]{#2}
  \newcommand{\nbcomment}[1]{}
   \newcommand{\atbcrcomment}[1]{}
    \newcommand{\agycomment}[1]{}
     \newcommand{\omcomment}[1]{}    
\else 
     \newcommand\nbcr[2]{\ifnum#1=\value{version}\textcolor{red}{#2}\else{\textcolor{blue}{#2}}\fi}
    \newcommand\atbcr[2]{\ifnum#1=\value{version}\textcolor{orange}{#2}\else{\textcolor{blue}{#2}}\fi}
    \newcommand{\atbcrcomment}[1]{\todo[size=\scriptsize, linecolor=orange, bordercolor=orange, backgroundcolor=white]{\textcolor{orange}{\textbf{@atb:} #1}}}
    \newcommand{\agycomment}[1]{\todo[size=\scriptsize, linecolor=purple, bordercolor=purple, backgroundcolor=white]{\textcolor{purple}{\textbf{@agy:} #1}}}

    \usepackage[colorinlistoftodos,prependcaption,textsize=small]{todonotes}
    \usepackage{marginnote} 
    \let\marginpar\marginnote
    \paperwidth=\dimexpr \paperwidth + 4cm\relax
    \oddsidemargin=\dimexpr\oddsidemargin + 2cm\relax
    \evensidemargin=\dimexpr\evensidemargin + 2cm\relax
    \marginparwidth=\dimexpr \marginparwidth + 2cm\relax
    \newcommand{\nbcomment}[1]{\todo[size=\scriptsize, linecolor=nbs, bordercolor=nbs, backgroundcolor=white]{\textcolor{nbs}{\textbf{@nb:} #1}}}
    \newcommand{\omcomment}[1]{\todo[size=\scriptsize, linecolor=omc, bordercolor=omc, backgroundcolor=white]{\textcolor{omc}{\textbf{@om:} #1}}}
\fi

\newcommand{\versionnum}[0]{2.1}

\newif\ifdraft
\draftfalse

\newif\ifrevision
\revisionfalse

\ifrevision
    \newcommand{\revcommon}[1]{\textcolor{cq}{#1}}
    \newcommand{\reva}[1]{\textcolor{ra}{#1}}
    \newcommand{\revb}[1]{\textcolor{rb}{#1}}
    \newcommand{\revc}[1]{\textcolor{rc}{#1}}

    \newcommand\revmid[2]{\todo[linecolor=#1,backgroundcolor=#1!15,bordercolor=#1]{\textcolor{black}{\textbf{#2}}}}

    \newcommand{\labela}[1]{\revmid{ra}{A#1}}
    \newcommand{\labelb}[1]{\revmid{rb}{B#1}}
    \newcommand{\labelc}[1]{\revmid{rc}{C#1}}

    \newcommand{\labelcq}[1]{\revmid{cq}{CQ#1}}

    \newcommand{\om}[1]{\textcolor{black}{#1}}

    \definecolor{babyblueeyes}{rgb}{0.63, 0.79, 0.95} 
    \usepackage[colorinlistoftodos,prependcaption,textsize=small]{todonotes}
    
\else
    \newcommand{\revcommon}[1]{{#1}}
    \newcommand{\reva}[1]{{#1}}
    \newcommand{\revb}[1]{{#1}}
    \newcommand{\revc}[1]{{#1}}

    \newcommand\revmid[2]{}

    \newcommand{\labela}[1]{\revmid{ra}{A#1}}
    \newcommand{\labelb}[1]{\revmid{rb}{B#1}}
    \newcommand{\labelc}[1]{\revmid{rc}{C#1}}

    \newcommand{\labelcq}[1]{\revmid{cq}{CQ#1}}

    \newcommand{\om}[1]{\textcolor{black}{#1}}

    \definecolor{babyblueeyes}{rgb}{0.63, 0.79, 0.95} 
\fi

\ifdraft
    \usepackage[colorinlistoftodos,prependcaption,textsize=scriptsize]{todonotes} % For margin notes
    \paperwidth=\dimexpr \paperwidth + 4cm\relax
    \oddsidemargin=\dimexpr\oddsidemargin + 2cm\relax
    \evensidemargin=\dimexpr\evensidemargin + 2cm\relax
    \marginparwidth=\dimexpr \marginparwidth + 2cm\relax
    \newcommand{\nbcomment}[1]{\todo[size=\scriptsize, linecolor=nbs, bordercolor=nbs, backgroundcolor=white]{\textcolor{nbs}{\textbf{@nb:} #1}}}
    \newcommand\param[1]{{\color{blue}{#1}}}
    \newcommand\nb[1]{{\color{nbs}{#1}}}
    
    \renewcommand{\om}[1]{\textcolor{omc}{#1}}

    \newcommand{\atbcomment}[1]{\todo[size=\scriptsize, linecolor=orange, bordercolor=orange, backgroundcolor=white]{\textcolor{orange}{\textbf{@atb:} #1}}}
    \newcommand\atb[1]{{\color{orange}{#1}}}
    
    \newcommand{\ouscomment}[1]{\todo[size=\scriptsize, linecolor=orange, bordercolor=orange, backgroundcolor=white]{\textcolor{ouscolor}{\textbf{@ous:} #1}}}
    \newcommand{\ous}[1]{\textcolor{ouscolor}{#1}}

    \newcommand{\ieycomment}[1]{\todo[size=\scriptsize, linecolor=orange, bordercolor=orange, backgroundcolor=white]{\textcolor{iy}{\textbf{@iey:} #1}}}
    
    \newcommand{\agy}[1]{\textcolor{gfored}{#1}}
    \newcommand{\agycomment}[1]{\todo[size=\scriptsize, linecolor=purple, bordercolor=purple, backgroundcolor=white]{\textcolor{purple}{\textbf{@agy:} #1}}}
    
    \newcommand{\outline}[1]{\textcolor{red}{\textbf{Outline:}~#1}}
    \newcommand{\inlinetodo}[1]{\textcolor{red}{\textbf{TODO:} #1}}
\else 
    \definecolor{shadecolor}{RGB}{180,180,180}
    \definecolor{darkgray}{rgb}{0.86, 0.86, 0.86}

    \newcommand\param[1]{{\color{black}{#1}}}

    \newcommand\nb[1]{#1}

    \renewcommand{\om}[1]{\textcolor{black}{#1}}

    \newcommand\atb[1]{#1}
    \newcommand{\atbcomment}[1]{}

    \newcommand{\ous}[1]{#1}
    \newcommand{\ouscomment}[1]{}

    \newcommand{\ieycomment}[1]{}
    
    \newcommand{\agy}[1]{#1}
    \renewcommand{\agycomment}[1]{}

    \newcommand{\outline}[1]{}
    \newcommand{\inlinetodo}[1]{}
\fi

\usepackage[framemethod=tikz]{mdframed}
\newcommand{\fancycommand}[1]{\begin{mdframed}[backgroundcolor=darkgray,linecolor=black,linewidth=0.7pt,skipabove=2pt]
    \texttt{#1}%
\end{mdframed}
\vspace{-\topsep}}

\usepackage{pgfornament}
\usepackage{tikz}
\usetikzlibrary{calc}

\makeatletter
\g@addto@macro{\normalsize}{%
  \setlength{\abovedisplayskip}{4pt plus 0.5pt minus 1pt}
  \setlength{\belowdisplayskip}{3pt plus 0.5pt minus 1pt}
  \setlength{\abovedisplayshortskip}{0pt}
  \setlength{\belowdisplayshortskip}{0pt}
  \setlength{\intextsep}{5pt plus 1pt minus 1pt}
  \setlength{\textfloatsep}{3pt plus 1pt minus 1pt}
  \setlength{\skip\footins}{5pt plus 1pt minus 1pt}
  \setlength{\abovecaptionskip}{2pt plus 0pt minus 0pt}}
\makeatother

%% file: ref_macros.tex
\newcommand{\exploitingRowHammerAllCitations}[0]{\cite{fournaris2017exploiting, poddebniak2018attacking, tatar2018throwhammer, carre2018openssl, barenghi2018software, zhang2018triggering, bhattacharya2018advanced, google-project-zero, kim2014flipping, rowhammergithub, seaborn2015exploiting, van2016drammer, gruss2016rowhammer, razavi2016flip, pessl2016drama, xiao2016one, bosman2016dedup, bhattacharya2016curious, burleson2016invited, qiao2016new, brasser2017can, jang2017sgx, aga2017good, mutlu2017rowhammer, tatar2018defeating, gruss2018another, lipp2018nethammer, van2018guardion, frigo2018grand, cojocar2019eccploit,  ji2019pinpoint, mutlu2019rowhammer, hong2019terminal, kwong2020rambleed, frigo2020trrespass, cojocar2020rowhammer, weissman2020jackhammer, zhang2020pthammer, yao2020deephammer, deridder2021smash, hassan2021utrr, jattke2022blacksmith, tol2022toward, kogler2022half, orosa2022spyhammer, zhang2022implicit, liu2022generating, cohen2022hammerscope, zheng2022trojvit, fahr2022frodo, tobah2022spechammer, rakin2022deepsteal, kang2024sledgehammer, lin2025gpuhammer,meyer2025phoenix,kamadan2025ecc}}

\newcommand{\understandingRowHammerNewCitations}[0]{\cite{kim2020revisiting, orosa2021deeper, orosa2022spyhammer, cohen2022hammerscope, yaglikci2022understanding, khan2018analysis, agarwal2018rowhammer, ni2018write, genssler2022reliability, mutlu2023fundamentally,olgun2025variable,tuugrul2025understanding,yuksel2025pudhammer,yuksel2025columndisturb}}

\newcommand{\mitigatingRowHammerAllCitations}[0]{\cite{AppleRefInc, rh-hp,rh-lenovo,greenfield2012throttling, kim2014flipping, kim2014architectural, bains14d, bains14c, bains14a, bains14b, aweke2016anvil, bains2015row, bains2016row, bains2016distributed, son2017making, seyedzadeh2018cbt,irazoqui2016mascat, you2019mrloc, lee2019twice, park2020graphene, yaglikci2021security, yaglikci2021blockhammer, frigo2020trrespass, kang2020cattwo, hassan2021utrr, qureshi2022hydra, saileshwar2022randomized, brasser2017can, konoth2018zebram, van2018guardion, vig2018rapid,  kim2022mithril, lee2021cryoguard, marazzi2022protrr, zhang2022softtrr, joardar2022learning, juffinger2023csi, yaglikci2022hira, saxena2022aqua, enomoto2022efficient, manzhosov2022revisiting, ajorpaz2022evax, naseredini2022alarm, joardar2022machine, hassan2022case, zhang2020leveraging,loughlin2021stop, devaux2021method, fakhrzadehgan2022safeguard, saroiu2022price, loughlin2022moesiprime, han2021surround, mutlu2023fundamentally, woo2023scalable, bock2019riprh, kim2015architectural, wang2021discreet, bennett2021panopticon, olgun2024abacus,bostanci2024comet,canpolat2024breakhammer,canpolat2025chronus,canpolat2024understanding,yaglikci2024svard,taneja2025dream,vittal2025mopac,lin2025cnc,qazi2025drfm,lu2025counterpoint,qureshi2025autorfm,woo2025dapper,woo2025qprac}}

\newcommand{\mitigatingRowHammerCounterCitations}[0]{\cite{kim2014flipping, seyedzadeh2018cbt, park2020graphene, qureshi2022hydra, kim2022mithril, marazzi2022protrr, yaglikci2022hira, loughlin2021stop, kim2014architectural,saxena2022aqua,saileshwar2022randomized,olgun2024abacus,bostanci2024comet}}

\newcommand{\refreshBasedRowHammerDefenseCitations}[0]{\cite{lee2019twice, seyedzadeh2017cbt, seyedzadeh2018cbt, kang2020cattwo, park2020graphene, kim2022mithril, kim2014architectural, bains2015row, bains2016distributed, bains2016row, aweke2016anvil,olgun2024abacus,bostanci2024comet,canpolat2025chronus}}

\newcommand{\rowHammerGetsWorseCitations}[0]{\cite{kim2014flipping, kim2020revisiting, frigo2020trrespass, yaglikci2022understanding, orosa2021deeper, mutlu2017rowhammer, mutlu2018rowhammer, mutlu2019rowhammer, mutlu2023fundamentally}}

\newcommand{\memorycontentionattacks}[0]{\cite{zhenyu2012whispers,dagli2025mc,zhao2022binoculars,liu2023side,chiang2025reload}}

%% file: glossary.tex
\newcommand{\nrh}[0]{N_{RH}}
\newcommand{\npr}[0]{N_{PR}}

\newcommand{\pprevref}[0]{p_{REF}}
\newcommand{\pnorefinaburst}[0]{p_{B}}
\newcommand{\cboost}[0]{c_B}
\newcommand{\creduc}[0]{c_R}
\newcommand{\pmin}[0]{p_{min}}
\newcommand{\pmax}[0]{p_{max}}

\newcommand{\pnorefuntilxy}[0]{P_{x,y}}
\newcommand{\pij}[0]{p_{i,j}}
\newcommand{\psa}[0]{p_{sa}}
\newcommand{\pfa}[0]{p_{fa}}
\newcommand{\pattack}[0]{p_{attack}}

\newcommand{\psecure}[0]{p_{security}}
\newcommand{\timeout}[0]{t_{to}}

\newcommand{\trcd}[0]{t_{RCD}}
\newcommand{\tras}[0]{t_{RAS}}
\newcommand{\trp}[0]{t_{RP}}
\newcommand{\trc}[0]{t_{RC}}
\newcommand{\trefi}[0]{t_{REFI}}
\newcommand{\trefw}[0]{t_{REFW}}
\newcommand{\trfc}[0]{t_{RFC}}
\newcommand{\trrd}[0]{t_{RRD}}

\newcommand{\act}[0]{ACT}
\newcommand{\pre}[0]{PRE}
\newcommand{\refresh}[0]{REF}
\newcommand{\writecmd}[0]{WR}
\newcommand{\rd}[0]{RD}

\newacronym{iqr}{$IQR$}{inter-quartile range}
\newacronym{act}{$\act{}$}{activate}
\newacronym{pre}{$\pre{}$}{precharge}
\newacronym{ref}{$\refresh{}$}{refresh}
\newacronym{wr}{$\writecmd{}$}{write}
\newacronym{rd}{$\rd{}$}{read}
\newacronym{nrh}{$\nrh$}{RowHammer threshold}
\newacronym{npr}{$\npr$}{preventive refresh threshold}
\newacronym{pprevref}{$\pprevref{}$}{preventive refresh probability}
\newacronym{pnorefinaburst}{$\pnorefinaburst$}{the probability of experiencing no preventive refresh during a burst}
\newacronym{pnorefuntilxy}{$\pnorefuntilxy$}{the probability of experiencing \emph{no} preventive refresh until a given $x_{th}$ row activation in $y_{th}$ burst of activations}
\newacronym{cboost}{$\cboost{}$}{the coefficient of boosting preventive refresh probability}
\newacronym{creduc}{$\creduc{}$}{the coefficient of reducing preventive refresh probability}
\newacronym{pij}{$\pij{}$}{the probability of experiencing no preventive refresh after the row activation $i$ at a given batch $j$}
\newacronym{psa}{$\psa{}$}{the probability of a successful attempt}
\newacronym{pfa}{$\pfa{}$}{the probability of a failed attempt}
\newacronym{pattack}{$\pattack{}$}{the probability of a successful attack}
\newacronym{psecure}{$\psecure{}$}{\gls{mech}'s security guarantee over a time period, T,}
\newacronym{trefw}{$\trefw$}{refresh window}
\newacronym{trcd}{$\trcd$}{{trcd definition here}}
\newacronym{tras}{$\tras$}{the latency of fully restoring a DRAM cell's charge}
\newacronym{trp}{$\trp$}{{trp definition here}}
\newacronym{trc}{$\trc$}{the minimum time needed between two consecutive row activations targeting the same bank}
\newacronym{trefi}{$\trefi$}{refresh interval}
\newacronym{trfc}{$\trfc$}{refresh latency}
\newacronym{trrd}{$\trrd$}{the minimum time needed between two consecutive row activations targeting the same rank}
\newacronym{timeout}{$\timeout$}{timeout period}
\newacronym{pmin}{$\pmin$}{{DEFINE PMIN PLEASE}}
\newacronym{pmax}{$\pmax$}{{DEFINE PMAX PLEASE}}

\newcommand{\rfmth}[0]{T_{RFM}}
\newcommand{\rfmab}[0]{RFM_{ab}}
\newcommand{\rfmsb}[0]{RFM_{sb}}
\newcommand{\aboth}[0]{N_{BO}}
\newcommand{\taboact}[0]{t_{ABO_{ACT}}}
\newcommand{\tbodelay}[0]{t_{BackOffDelay}}

\newcommand{\pmitigth}[0]{T_{FRRFM}}
\newcommand{\pmitig}[0]{FR-RFM}
\newcommand{\pmitiglong}[0]{Fixed-Rate RFM}
\newcommand{\pracrand}[0]{PRAC-RIAC}

\newacronym{rfm}{RFM}{refresh management}
\newacronym{prfm}{PRFM}{Periodic RFM}
\newacronym{rfmth}{$\rfmth{}$}{\emph{bank activation threshold}}
\newacronym{rfmab}{$\rfmab{}$}{all bank refresh management}
\newacronym{rfmsb}{$\rfmsb{}$}{same bank refresh management}
\newacronym{prac}{PRAC}{Per Row Activation Counting}
\newacronym{abo}{$ABO$}{Alert Back-Off}
\newacronym{aboth}{$\aboth{}$}{the back-off threshold}
\newacronym{taboact}{$\taboact{}$}{\emph{the window of normal traffic}}
\newacronym{tbodelay}{$\tbodelay{}$}{the delay until a new back-off can be initiated}
\newacronym{jedec}{JEDEC}{Joint Electron Device Engineering Council}

\newacronym{pmitigth}{$\pmitigth{}$}{the RFM period of \pmitig{}}
\newacronym{pmitig}{\pmitig}{Fixed-Rate RFM}

%% file: sections/00_abstract.tex
DRAM chips are increasingly vulnerable to read disturbance phenomena (e.g., RowHammer and RowPress), where repeatedly accessing {or keeping open} a DRAM row causes bitflips in nearby rows, due to DRAM density scaling. Attackers can \nbcr{1}{exploit} \rh{} bitflips in real systems \nbcr{1}{to compromise security, which has motivated many prior works on \rh{} defenses.}
\nbcr{1}{To enable such defenses,} recent DDR specifications 
\nbcr{1}{introduce} 
new \nbcr{1}{defense} frameworks (e.g., PRAC and RFM).
For \om{robust (i.e., secure, safe, and reliable) operation}, it is critical to analyze security implications \nbcr{1}{of} widely-adopted \rh{} \nbcr{1}{defenses}.
{\nbcr{1}{Yet}, \textit{no} prior work analyzes the timing covert channel and side channel vulnerabilities \nbcr{1}{\nbcr{2}{\rh{} defenses} introduce.}}

{\om{This paper presents} the first analysis and evaluation of timing covert \nbcr{1}{channel} and side channel vulnerabilities introduced by \om{state-of-the-art} \rh{} \nbcr{1}{defenses}.}
{\om{We demonstrate} that} \rh{} \nbcr{1}{defenses}' \emph{preventive actions} (e.g., preventively refreshing potential victim rows)
{have two \om{fundamental} features {that} {allow an attacker to exploit
\rh{} \nbcr{1}{defenses} for \nbcr{1}{timing} 
\nbcr{1}{leakage}.} First,} {preventive actions often reduce DRAM bandwidth availability because they block access to DRAM, thereby 
resulting in significantly \om{longer} memory access latencies}.
{Second, \nbcr{0}{users can \nbcr{1}{intentionally} trigger preventive actions because preventive actions highly depend on \nbcr{1}{application} memory access patterns.}}

\nb{We 
introduce} \X{}, a new class of attacks that leverage the {\rh{}} \nbcr{1}{defense}-induced memory latency differences to establish communication channels between processes and leak secrets from victim processes.  First, we build {\param{two}} covert channel attacks exploiting {\param{two}} {state-of-the-art \rh{} \nbcr{1}{defenses} \nbcr{0}{(i.e., PRAC and RFM)}, \nb{achieving}
\param{39.0 Kbps} and \param{48.7 Kbps} {channel capacity}.}
Second, we demonstrate a proof-of-concept 
website fingerprinting attack that can identify visited websites based on the \nb{\rh{}-preventive actions they cause.} 
{We propose and evaluate \nbcr{0}{{three}} countermeasures against \X{}. \nbcr{1}{Our results} show that {fundamentally \nbcr{2}{and completely} mitigating} \X{} {induces} \om{large} performance overheads in highly \rh{}-vulnerable systems.
We believe and hope our work can enable and aid future work on designing \nbcr{2}{better solutions and more} robust systems \nbcr{2}{in the presence of such new vulnerabilities.}}

%% file: sections/01_intro.tex
\section{Introduction}
\label{sec:intro}
{DRAM chips are susceptible to read disturbance where repeatedly accessing} \om{or keeping open} {a DRAM row} (i.e., \textit{an aggressor row}) can cause bitflips in physically nearby rows (i.e., \textit{victim rows})~\cite{kim2014flipping, mutlu2017rowhammer, yang2019trap, mutlu2019rowhammer,park2016statistical, park2016experiments,
walker2021ondramrowhammer, ryu2017overcoming, yang2016suppression, yang2017scanning, gautam2019row, jiang2021quantifying,luo2023rowpress,mutlu2023fundamentally}.
{RowHammer~\cite{kim2014flipping,mutlu2023fundamentally,mutlu2019rowhammer,mutlu2017rowhammer} is a type of read disturbance phenomenon, \nb{where a victim row can experience bitflips when at least one nearby row is hammered more times than a threshold, called the \gls{nrh}.}
Modern DRAM chips become more vulnerable to \rh{} as DRAM technology {node size becomes smaller}~\rowHammerGetsWorseCitations{}. 
\emph{RowPress}~\cite{luo2023rowpress} is another example DRAM read disturbance \om{phenomenon} that amplifies \nb{\om{bitflips by} keeping the aggressor row open for
longer, thereby causing more disturbance with each activation.} 
Prior works show that {attackers can leverage} \rh{} bitflips in real systems~\exploitingRowHammerAllCitations{} to{, for example,} (i)~take over \nbcr{1}{an otherwise secure} system by {escalating privilege}, (ii) leak security-critical and private data \nbcr{1}{(e.g., cryptographic keys machine learning model parameters)}, \nbcr{1}{(iii)~crash a system, (iv)~render machine learning inference inaccurate by corrupting important data, and (v) break out of virtual machine sandboxes}. \nbcr{1}{To avoid such problems}, many prior works from academia and industry propose \om{various} \nbcr{1}{defenses} to prevent \rh{} bitflips~\mitigatingRowHammerAllCitations{}. \om{Recent} \om{DDR5} specifications \om{introduce} new \nbcr{1}{defense} frameworks such as \gls{prac}~\cite{jedec2024jesd795c,canpolat2024understanding} \nb{(as of April 2024)} and \gls{rfm}~\cite{jedec2020jesd795} \nb{before 2024)}.
\om{The industry \nbcr{1}{has} already adopted multiple \rh{} \nbcr{1}{defenses}, yet their full security implications are not known. In this work, we ask the question: \emph{Do industrial and academic \rh{} \nbcr{1}{defenses} introduce new covert and side channel vulnerabilities?} Unfortunately, the answer is \emph{yes}.}

\nb{This work presents} the first analysis and evaluation of timing covert \nbcr{1}{channel} and side channel vulnerabilities introduced by \om{state-of-the-art} \rh{} \nbcr{1}{defenses}.
{Our key observation is that} \rh{} \nbcr{1}{defenses}' \emph{preventive actions} (e.g., preventively refreshing potential victim rows, migrating aggressor rows, throttling accesses \nbcr{1}{to} frequently-accessed rows) {have two \om{fundamental} features \om{that} {allow an attacker to exploit
\rh{} \nbcr{1}{defenses} for covert and side channels.} First, {preventive actions}}
\nbcr{0}{often reduce DRAM bandwidth availability because they block access to DRAM for regular memory requests (e.g., by preventively refreshing potential victim rows \nbcr{1}{or} creating off-chip movement of aggressor rows' content or metadata).}
{Second, \nbcr{0}{a user can \nbcr{1}{intentionally} trigger a preventive action because preventive actions highly depend on \nbcr{1}{application} memory access patterns.}}

{We systematically analyze the \param{two} 
latest industry \nbcr{1}{defenses} \nbcr{1}{(PRAC~\cite{jedec2024jesd795c,canpolat2024understanding} and RFM~\cite{jedec2020jesd795})} to {\rh{}}\footnote{We qualitatively analyze other \nbcr{1}{\rh{}} \nbcr{1}{defenses} proposed by academia and industry in \secref{sec:other-mechanisms}.} and %
introduce} \emph{\X{}}, a new class of attacks that leverage the {\rh{}} \nbcr{1}{defense}-induced memory latency differences to establish communication channels between processes and leak secrets from victim processes. \nbcr{0}{\X{}'s key idea is to exploit \rh{} \nbcr{1}{defenses} to 1) deterministically \nbcr{1}{and intentionally} impose high latency on memory requests via preventive actions and 2) accurately infer other applications' memory access patterns that result in preventive actions.}

\head{\X{} Covert Channels} We build \atb{\param{two}} covert channel attacks exploiting \atb{\param{two}} state-of-the-art \rh{} \nbcr{1}{defenses}: \gls{prac}~\cite{jedec2024jesd795c,canpolat2024understanding} and \gls{prfm}~\cite{jedec2020jesd795}. These covert channels transmit messages between sender and receiver processes \nbcr{0}{by encoding messages \nbcr{1}{using} the preventive actions of \nbcr{2}{the defense} mechanisms ({\nbcr{0}{i.e., PRAC \nb{back-offs} and \gls{rfm} commands}). The receiver decodes \nbcr{1}{a} message by detecting preventive actions via measuring \nbcr{1}{the} latencies of its own memory requests. Using this method,} our covert channels {provide} \param{39.0 Kbps} and \param{48.7 Kbps} \nb{average} {channel capacity}.\footnote{\nbcr{0}{\secref{sec:methodology}} describes our evaluation methodology in detail.}}

\head{\X{} Side Channel} We demonstrate a proof-of-concept \nb{website fingerprinting attack}
that \nbcr{0}{identifies \nbcr{1}{the} website  visited by a victim user.} 
{\nbcr{0}{Our attack \nbcr{1}{works} by} observing \nb{preventive actions \nbcr{0}{that the browser \nbcr{1}{triggers}} in a system \nbcr{1}{employing} \gls{prac}} during website \nbcr{1}{loading} \nbcr{0}{and \nbcr{1}{constructs} a \textit{fingerprint}. Our evaluations show that \nbcr{1}{the} attacker can \nbcr{1}{classify the victim's website with high accuracy} based on \nbcr{1}{the fingerprint} of preventive actions.}}

\head{\X{} Countermeasures}
We propose and evaluate \nbcr{0}{three} countermeasures against \X{}:
\nbcr{1}{1)~\textit{\gls{pmitig}}, which decouples preventive actions from application memory access patterns} by performing preventive actions at a fixed rate, thereby completely \nbcr{1}{eliminating the timing channel, 2)}~\emph{Randomly Initialized Activation Counters (RIAC)}, which reduces \X{}'s channel capacity by initializing activation counters with random values after each preventive action, \nbcr{1}{thereby introducing unintentional preventive actions at random activation counts and reducing the reliability of the channel, 3)}~\emph{Bank-Level PRAC}, which performs preventive actions \nbcr{2}{separately for each} bank, thereby \nbcr{1}{preventing an attacker from observing preventive actions across multiple banks and} reducing \X{}'s scope to the level of existing DRAM-based attacks.

\nbcr{1}{Based on our evaluations, we make three key observations. First, at a near-future \gls{nrh} value of 1024, \gls{pmitig} mitigates \X{} and \rh{} while performing similarly to \gls{prac} and \gls{rfm}, which are insecure against \X{}. Second, at very low \gls{nrh} values (e.g., $\le$128), \nbcr{1}{\gls{pmitig}} incurs high performance overheads. At these \gls{nrh} values, RIAC reduces the channel capacity and induces a lower performance overhead (e.g., $2.14\times$ at \gls{nrh}=64) compared to \gls{pmitig} (e.g., $18.2\times$ at \gls{nrh}=64). Third, Bank-Level PRAC prevents attacks that observe preventive actions across different banks at all \gls{nrh} values. However, it does not mitigate \X{} within a DRAM bank.}

We conclude that fundamentally mitigating \X{} \nbcr{1}{with \gls{pmitig}} incurs \nbcr{0}{a low performance overhead at near-future \gls{nrh} values,} and high performance overheads \nbcr{0}{at very low \gls{nrh} values} \nb{(e.g., \gls{nrh}$\le$128)} and requires further research. We believe and hope our work can \nbcr{1}{inspire and} enable future work on designing \nbcr{2}{better solutions and more} robust systems \nbcr{2}{in the presence of such new vulnerabilities}.}

This work makes the following contributions:
\begin{itemize}
\item  \nbcr{2}{For the first time,} \om{we demonstrate} timing \nbcr{1}{covert channels and side channels}  
introduced by \rh{} \nbcr{1}{defenses}.
\item We present \X{}, a new class of attacks that leverage the \rh{} \nbcr{1}{defense}-induced memory latency differences to establish \nbcr{2}{new} \nbcr{1}{covert channels and side channels}. 
\item We \nb{demonstrate} two \X{} covert channel attacks exploiting two state-of-the-art \rh{} \nbcr{1}{defenses \nbcr{2}{used in industry}}. Our covert channels maintain their channel capacity while running concurrently with memory-intensive applications.
\item We demonstrate a \X{} side channel: a website fingerprinting attack that can identify \om{which} websites \om{are} loaded in a web browser. We showcase that classical machine learning models can learn to distinguish websites based on the characteristics of the \om{\rh{}-}preventive actions they trigger.
\item We propose and evaluate \nbcr{0}{three} new countermeasures against \X{}. \nbcr{0}{We show that for near future \gls{nrh} values (e.g., 1024), \nbcr{1}{one of our countermeasures, \gls{pmitig}, completely eliminates \X{} timing channel with} low performance overheads.
\nbcr{2}{In systems that are highly \rh{}-vulnerable,} \nbcr{1}{\gls{pmitig}'s} performance overhead increases, \nbcr{1}{making countermeasures that reduce the timing channel capacity more practical due to their lower performance overhead}.} 
\item \nbcr{1}{To enable future research, we open source our simulation infrastructure and results at \url{https://github.com/CMU-SAFARI/LeakyHammer}.}
\end{itemize}

%% file: sections/02_background.tex
\section{Background}

\subsection{DRAM Organization and Operation}

\begin{figure}[ht]
\centering
\includegraphics[width=0.85\linewidth]{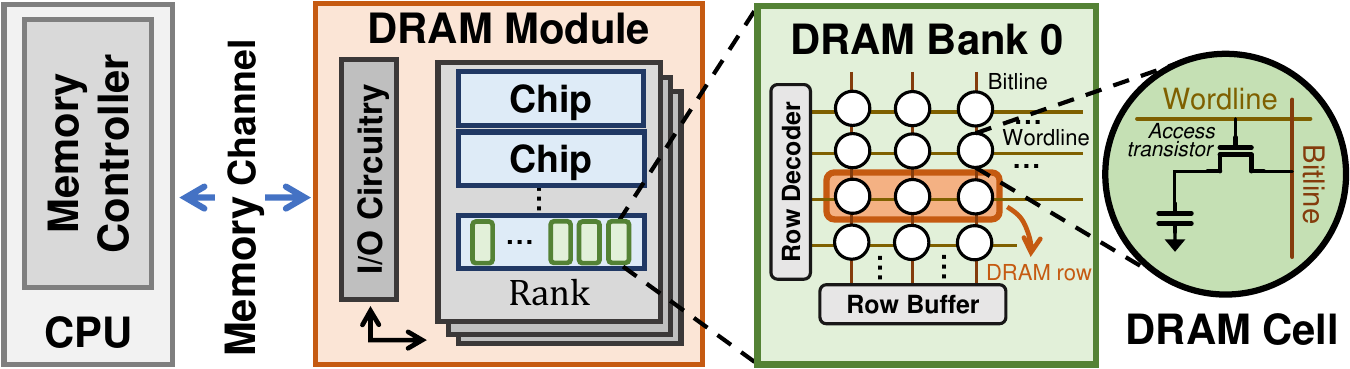}
\caption{DRAM organization.}
\label{fig:dram-organization}
\end{figure}

\head{Organization}~\figref{fig:dram-organization} shows the hierarchical organization of modern DRAM-based main memory. The memory controller connects to a DRAM module over a memory channel. A module contains one or {more} DRAM ranks that time-share the memory channel. A rank consists of multiple DRAM chips that operate in lock-step. 
Each DRAM chip contains multiple DRAM banks that can be accessed independently. A DRAM bank is organized as a two-dimensional array of DRAM cells, where a row of cells is called \textit{a DRAM row}. A DRAM cell consists of 1) a storage capacitor, which stores one bit of information in the form of electrical charge, and 2) an access transistor, which connects the capacitor to the row buffer through a bitline controlled by a wordline.

\head{Operation}
To access a DRAM row, the memory controller issues a set of commands to DRAM over the memory channel. The memory controller sends an \gls{act} command to activate a DRAM row, which asserts the corresponding wordline and loads the row data into the row buffer. Then, the memory controller can issue $RD$/$WR$ command{s} to read from/write into the DRAM row.
Subsequent accesses to the same row cause a row hit. To access a different row, the memory controller must first close the bank by issuing a \gls{pre} command.
Therefore, accessing a different row causes a row buffer miss/conflict.

DRAM cells are inherently leaky and lose their charge over time due to charge leakage in the access transistor and the storage capacitor~\cite{liu2012raidr,liu2013experimental}. {T}o maintain data integrity, the memory controller periodically refreshes each row in a time interval called \gls{trefw} {which is typically} 32~$ms$ for DDR5~\cite{jedec2020jesd795} and 64~$ms$ for DDR4~\cite{jedec2017ddr4}. 
To ensure all rows are refreshed every \gls{trefw}, the memory controller issues REF commands with a time interval called \gls{trefi} ($3.9~\mu s$ for DDR5~\cite{jedec2020jesd795} and $7.8~\mu s$ for DDR4~\cite{jedec2017ddr4} {at normal operating temperature {range}}).

\head{Timing Parameters}
To ensure correct operation, the memory controller must obey specific timing parameters while accessing DRAM~\cite{salp,lee2013tiered}. In addition to \gls{trefw} and \gls{trefi}, we explain \param{three} timing parameters related to the rest of the paper: i) \gls{trc}, ii) \gls{tras}, and iii) the minimum time needed to issue an ACT command following a PRE command ($t_{RP}$).

\subsection{DRAM Read Disturbance}
As DRAM manufacturing technology node size \nbcr{1}{reduces}, interference {between cells} increases, causing \nbcr{1}{device-level and} circuit-level read disturbance mechanisms~\understandingRowHammerNewCitations{}.
Two prime examples of such read disturbance mechanisms are RowHammer~\cite{kim2014flipping,mutlu2023fundamentally,mutlu2019rowhammer,mutlu2017rowhammer,mutlu2023retrospective} and RowPress~\cite{luo2023rowpress,luo2024rowpress}, where repeatedly activating 
%(i.e., opening) 
a DRAM row (i.e., aggressor row) or keeping the aggressor row active for a long time induces bitflips in physically nearby rows (i.e., victim rows), respectively.
To induce RowHammer bitflips, an aggressor row needs to be activated more {times} than a threshold value called \gls{nrh}. 

\head{DRAM Read Disturbance \nbcr{1}{Defenses}}
Many prior works propose \nbcr{1}{defenses}~\mitigatingRowHammerAllCitations{} to protect DRAM chips against RowHammer bitflips.
These {\nbcr{1}{defenses} usually} perform two tasks: 1)~execute a trigger algorithm and 2)~perform preventive actions.
The {\emph{trigger algorithm}} observes memory access patterns and triggers a {\emph{preventive action}} based on the result of a probabilistic or a deterministic process.
Preventive actions include 1)~preventively refreshing victim \nbcr{1}{rows}~\refreshBasedRowHammerDefenseCitations{},
2)~dynamically remapping aggressor rows~\cite{saileshwar2022randomized, saxena2022aqua, wi2023shadow, woo2023scalable}, and
3)~throttling unsafe accesses~\cite{greenfield2012throttling, yaglikci2021blockhammer,canpolat2024breakhammer}.
Existing RowHammer \nbcr{1}{defenses} can also prevent RowPress bitflips when they are configured for 
lower \gls{nrh} values~\cite{luo2023rowpress}.

\section{Motivation}
\label{sec:motivation}
\nbcr{1}{Cell density} scaling~\cite{meza2013case, mutlu2013memory, chang2016understanding, chang2017understanding, ghose2019demystifying,mandelman2002challenges,mutlu2025memory} exacerbates DRAM cells' vulnerability to read disturbance phenomena, where accessing \nb{and keeping open} a DRAM cell disturbs and can cause a bitflip in a physically nearby \emph{unaccessed} cell~\cite{redeker2002investigation, kim2014flipping, kim2020revisiting, luo2023rowpress}. RowHammer~\cite{kim2014flipping} and RowPress~\cite{luo2023rowpress} are two prime examples of DRAM read disturbance \nbcr{1}{that get exacerbated with} technology node scaling \nbcr{1}{which} makes them more prominent challenges going forward~\cite{kim2020revisiting, luo2023rowpress,mutlu2025memory,mutlu2019rowhammer,mutlu2017rowhammer,mutlu2023fundamentally}. To ensure robust (i.e., secure, safe, and reliable) \nb{operation} in \nb{current} and future systems, many prior works~\mitigatingRowHammerAllCitations{} propose \nb{various} RowHammer \nbcr{1}{defenses}, and their adaptations to the RowPress phenomenon~\cite{luo2023rowpress, saxena2024impress, yauglikcci2024spatial, olgun2024abacus}. 
\om{Recent} \om{DDR5} specifications \om{introduce} new \nbcr{1}{defense} frameworks such as \gls{prac}~\cite{jedec2024jesd795c,canpolat2024understanding} \nb{(as of April 2024)} and \gls{rfm}~\cite{jedec2020jesd795} \nb{(before 2024)}.
\nb{Industry \nbcr{1}{has} already adopted multiple \rh{} \nbcr{1}{defenses;} yet their full security implications are not known. In this work,} 
\nb{our goal is to analyze the covert \nbcr{1}{channel} and side channel vulnerabilities \nbcr{1}{due to} industrial and academic \rh{} \nbcr{1}{defenses}. }

%% file: sections/03_rhsc.tex
\section{\X{}: \rh{} Defense-Based Timing Attacks}

The key observation that enables \X{} is that \rh{} defenses' {preventive actions} (e.g., preventively refreshing potential victim rows)
\agy{have two \nb{fundamental} features that allow an attacker to exploit
\rh{} defenses for covert and side channels.}

First, \nb{preventive actions} often result in \nb{high} memory access latencies for regular memory requests as they either 1) \nbcr{1}{cause} DRAM \nbcr{1}{to be} unavailable for specific time intervals due to refreshing a set of DRAM rows~\refreshBasedRowHammerDefenseCitations{} or 2) create contention in DRAM \agy{due to off-chip data movement for migrating DRAM rows' contents or \rh{} tracking metadata~\cite{saileshwar2022randomized, wi2023shadow, woo2023scalable, saxena2022aqua, qureshi2022hydra}}. \nbcr{1}{A preventive} action is often a costly operation that can be observed from \nbcr{0}{userspace} applications (as we show in~\secref{sec:observe_prac} and~\secref{sec:rfm_characterization}), with a latency in the order of several hundred nanoseconds \ous{(e.g., 1400 ns for PRAC~\cite{jedec2024jesd795c,canpolat2024understanding})}.

\agy{Second, a user can \nbcr{1}{intentionally} trigger a preventive action because preventive actions highly depend on memory access patterns.}
Secure \rh{} defenses deterministically perform preventive actions based on their trigger algorithms to ensure that no DRAM row is activated enough times to cause a \rh{} bitflip. \nbcr{1}{A user} can activate the same row \gls{nrh} times to deterministically trigger \atb{a} \rh{} defense to perform a preventive action. 

Based on this key observation, we introduce \X{}, a new class of attacks that leverage the \rh{} defense-induced memory latency differences to establish \nbcr{1}{covert} communication channels \nbcr{1}{\nbcr{2}{between} attack processes} and leak \nbcr{1}{sensitive} data. The key idea of \X{} is to exploit \rh{} defenses to 1) deterministically \nbcr{1}{and intentionally} impose high latency on other requests via preventive actions and 2) accurately infer other applications' memory access patterns that result in preventive actions.

%% file: sections/04_methodology.tex
\section{Methodology}
\label{sec:methodology}

\subsection{Modeling Near-Future Systems with Secure \rh{} \nbcr{1}{Defenses}}

\atb{PRAC~\cite{jedec2024jesd795c,canpolat2024understanding} and RFM~\cite{jedec2020jesd795} are two key \rh{} defenses with strong potential
to securely mitigate \rh{} in near-future systems. These mechanisms are
already incorporated into the existing DDR5 standard~\cite{jedec2020jesd795}, but both are relatively new. Therefore, they are \textit{not yet} widely adopted but will be essential for future memory systems. {Our paper performs the best practice by simulating a variety of future systems.}
}

We faithfully model a \atb{high-performance computing} system with a system simulation platform, gem5~\cite{gem5}. We integrate gem5 with a cycle-level DRAM simulator, \nbcr{1}{Ramulator 2.0}~\cite{kim2016ramulator, luo2023ramulator2, ramulator2github}, to model the memory system with different \rh{} \nbcr{1}{defenses}. Table~\ref{tab:configs} shows the system and memory configurations that we use in our evaluation. 

\input{method_table}

\head{{Real System Noise}}
\revc{We faithfully and rigorously model various noise sources 1) periodic refreshes~\cite{liu2012raidr}, 2) real-world application-induced interference~\cite{mutlu2008parbs,subramanian2014bliss,subramanian2016bliss}, 3) memory traffic caused by data prefetchers, and \nbcr{1}{4) \nbcr{2}{additional} \rh{} preventive actions \nbcr{2}{caused by \rh{} defenses} (explained in~\secref{sec:c1_high})}. We evaluate LeakyHammer with a wide range of noise levels.
}

\subsection{Threat Model and Metrics}
\label{sec:threat_model}
\head{Covert Channel} We assume a scenario where a sender and receiver execute on the same system to exchange information. Both processes \nbcr{1}{can access} 1) fine-grained timers, such as the \texttt{rdtsc} instruction~\cite{intelmanual2016}, and 2) instructions to flush cache blocks, such as the \texttt{clflush} and \texttt{clflushopt} instructions~\cite{intelmanual2016}. \nbcr{1}{In our evaluation, we obtain timestamps using m5 operations~\cite{gem5} and simulate the \texttt{clflush} instruction and its effects in the simulation environment.}

{The sender and receiver can observe timing differences caused by the \rh{} \nbcr{1}{defense} by allocating different rows in the same channel for PRAC~\cite{jedec2024jesd795c,canpolat2024understanding} and in the same bank across all bank groups for RFM~\cite{jedec2020jesd795}. This is because \nbcr{1}{PRAC blocks all accesses \nbcr{2}{to an entire} channel, while RFM blocks accesses to the same bank in each bank group (and in some cases all banks \nbcr{2}{in the bank group})}. In our attacks, we colocate sender and receiver data in the same bank to cause row activations with \nbcr{1}{row buffer conflicts} (explained in more detail in~\secref{sec:case_one} and~\secref{sec:case_two}). \nbcr{1}{However, \nbcr{1}{data} colocation at the bank level} is not \nbcr{1}{required:} 
\nbcr{1}{attack processes} can \nbcr{1}{independently} access \nbcr{1}{two rows within a bank to create row buffer conflicts}.}

Locating pages in the same bank \textit{does not} require sharing actual data (i.e., having access to the same page).
Sender and receiver processes can place their pages intentionally in the same bank by \nbcr{1}{partially} reverse engineering the address mapping (e.g., via reverse engineering tools proposed in prior works~\cite{pessl2016drama,zenhammer}) and using memory massaging techniques~\cite{kwong2020rambleed,pessl2016drama,van2016drammer,dramaqueen}. 

We evaluate our covert channels using the \textit{channel capacity} metric~\cite{shusterman2019robust}. Channel capacity is the raw bit rate \nbcr{1}{(i.e., number of transmitted bits per second)} multiplied with $1 -H(e)$, where $e$ is the error probability \nbcr{1}{(i.e., number of erroneous bits over the total number of bits transmitted)} and $H(e)$ is the binary entropy function. It is calculated as follows.
\vspace{-0.75mm}
\begin{equation}
\begin{split}
Channel Capacity = Raw Bit Rate \times \left( 1 - H(e) \right)
\\
\quad H(e) = -e \log_2(e) - (1-e) \log_2(1-e)
\end{split}
\end{equation}

\head{PRAC-Based Side Channel} The attacker process executes on the same system as the victim process and has the same abilities as the sender and receiver processes. The attacker can observe the preventive actions caused by other applications accessing the same channel. We describe the threat model and metrics in more detail in \secref{sec:sidechannel}.

%% file: method_table.tex
\begin{table}[h]
\caption{Evaluated System Configurations.}
\resizebox{\linewidth}{!}{%
\begin{tabular}{ll}
\hline
\multicolumn{2}{c}{\textbf{gem5: System Configuration}} \\ \hline
\textbf{Processor}: &
  x86, 1-,2-,4-core, out-of-order, \SI{3}{\giga\hertz} \\
\textbf{L1 Data + Inst. Cache:} &
  \SI{32}{\kilo\byte}, 8-way, \SI{64}{\byte} cache line \\
\textbf{Last-Level Cache:} &
  \SI{4}{\mega\byte} \revcommon{(per core)}, 16-way, \SI{64}{\byte} cache line \\ \hline
\multicolumn{2}{c}{\textbf{Ramulator 2.0: Memory Controller \& Main Memory}} \\ \hline
\multirow{3}{*}{\textbf{Memory Controller:}} &
  \multirow{3}{*}{\begin{tabular}[c]{@{}l@{}}64-entry read and write request queues,\\ Scheduling policy: FR-FCFS~\cite{rixner2000memory,zuravleff1997controller}\\ with a column cap of 16~\cite{mutlu2007stall}\end{tabular}} \\
 &
   \\
 &
   \\
\multirow{2}{*}{\textbf{Main Memory:}} &
  \multirow{2}{*}{\begin{tabular}[c]{@{}l@{}}DDR5, 1 channel, 2 rank/channel, 8 bank groups,\\ 4 banks/bank group, 128K rows/bank\end{tabular}} \\
 &
   \\ \hline
\end{tabular}
}
\label{tab:configs}
\end{table}

%% file: sections/05_0_casestudy_prac.tex
\section{Case Study 1: PRAC-based Covert Channel}
\label{sec:case_one}
\subsection{\agy{PRAC-based RowHammer Defense}}

\head{RFM Command}
\agy{\Gls{rfm}} is a DRAM command that provides the DRAM chip with a time window 
{(\agy{e.g.,} \param{\SI{350}{\nano\second}}~\cite{jedec2024jesd795c})}
\agy{to preventively refresh} potential victim rows.
\agy{The DRAM chip identifies potential victim rows and the memory controller issues \gls{rfm} commands.} \nbcr{1}{See \cite{canpolat2024understanding,canpolat2025chronus} for more detail.}

\head{\gls{prac} Overview}
\agy{\gls{prac} accurately measures each DRAM row's activation count by implementing a counter per row.}
When a row's activation count reaches a threshold, the DRAM chip \agy{needs to dedicate a time window for preventively refreshing potential victim rows. As the memory controller has fine-grained control over the DRAM operations and timings, the DRAM chip notifies the memory controller by \nbcr{1}{asserting the} alert-back-off (ABO) \nbcr{1}{signal}. ABO} forces the memory controller to issue an RFM command \agy{soon enough so that the DRAM chip safely performs preventive refreshes} 
upon receiving \agy{the} RFM command.

\head{\gls{prac}'s Operation and Parameters}
\gls{prac} increments a DRAM row's activation count while the row is being closed. 
The DRAM chip asserts the back-off signal when a row's activation count reaches a fraction \agy{(e.g., 70\%, 80\%, 90\%, or 100\%~\cite{jedec2024jesd795c})} of \gls{nrh}, denoted as \gls{aboth}.
The memory controller receives the back-off signal \agy{shortly (e.g., \param{$\approx$\SI{5}{\nano\second}}~\cite{jedec2024jesd795c}) after issuing a \gls{pre} command.}
\agy{Then, the memory controller serves requests normally for a limited time window called \gls{taboact} (e.g., \param{\SI{180}{\nano\second}})~\cite{jedec2024jesd795c}.}
\agy{At the end of \gls{taboact}, the DRAM chip undergoes a recovery period, where the memory controller issues a number of \gls{rfm} commands (e.g., 1, 2, or 4)~\cite{jedec2024jesd795c}, and thus the DRAM chip refreshes potential victim rows around the rows with the highest activation counts.} An \gls{rfm} command can further increment the activation count of a row before its potential victims are refreshed.
\agy{The DRAM chip needs to respect a cool-down window~\cite{jedec2024jesd795c} before asserting the ABO signal again after the recovery period, during which several row activation operations can be performed.}

\nbcr{1}{\head{Assumptions about PRAC's Implementation}
We make the following \param{two} assumptions similar to prior works~\cite{canpolat2024understanding,canpolat2025chronus}. First,
the memory controller issues \agy{four} back-to-back \gls{rfm} commands after receiving a back-off signal. \nb{Thus, the DRAM chip refreshes four potential aggressor rows' victims during one back-off.}
\agy{Second, \gls{aboth} is \nbcr{2}{set to 128. Doing so} (i)~alleviates significant testing overhead \nbcr{2}{by testing DRAM chips only for 128 activations to ensure no \rh{} bitflips occur,} and (ii)~keeps \gls{prac}'s performance overhead low~\cite{canpolat2025chronus,canpolat2024understanding}.}}

\subsection{PRAC-induced Memory \agy{Access Latency}}
\label{sec:observe_prac}
To observe the PRAC-induced latencies from userspace applications, we construct a routine that 1)~triggers back-offs with a memory access loop and 2)~measures the memory request latencies within the loop.  \agy{Listing~\ref{ls:latency_measure} shows our routine.}
First, to trigger PRAC back-offs, the routine 
\nbcr{1}{allocates pages in \agy{two DRAM rows within a bank} (line 4) and accesses them} in an \nbcr{1}{interleaved} manner to create row buffer conflicts \agy{to increase the activation counts of these two rows} \agy{(line~13)}.
Second, the routine measures the access loop's \agy{execution} time, \agy{and thus captures high-latency events, including periodic refreshes \agy{(lines 8-17)}.}  
\agy{To decrease the chance of missing periodic refreshes scheduled in \agy{an unobserved} time window, we measure the execution time \nb{continuously}} \agy{similar to \agy{a} prior work's \agy{methodology}~\cite{zenhammer}} \nb{and use the timestamp (line~15) in the access loop as} 1)~the end time of the current iteration \agy{(line~17)} and 2)~the start time of the next iteration \agy{(line~18)}.

\begin{figure}[h]
\begin{lstlisting}[style=customc, label = ls:latency_measure, caption = Memory request latency measurement routine.]
 // row_ptrs array has two address pointers
 // located in separate DRAM rows
 // in the same DRAM bank
 vector<uint64_t> measure (vector<char*>& row_ptrs) {
    vector<uint64_t> measured_latency(ITERATIONS, 0);
    // get start timestamp
    uint64_t start = m5_rpns(); 
    for (int i = 0; i < ITERATIONS; i++) {
        int a = i % row_ptrs.size();
        auto row_ptr = row_ptrs[a];
        clflush(row_ptr); 
        // access a target row
        *(volatile char*)row_ptr;
        // get end timestamp
        uint64_t end  = m5_rpns();
        // record the measured latency 
        measured_latency[i] = end-start;
        start = end;
    }
    return measured_latency;
}
\end{lstlisting}
\end{figure}

\begin{figure}[h]
\centering
\includegraphics[width=0.95\linewidth]{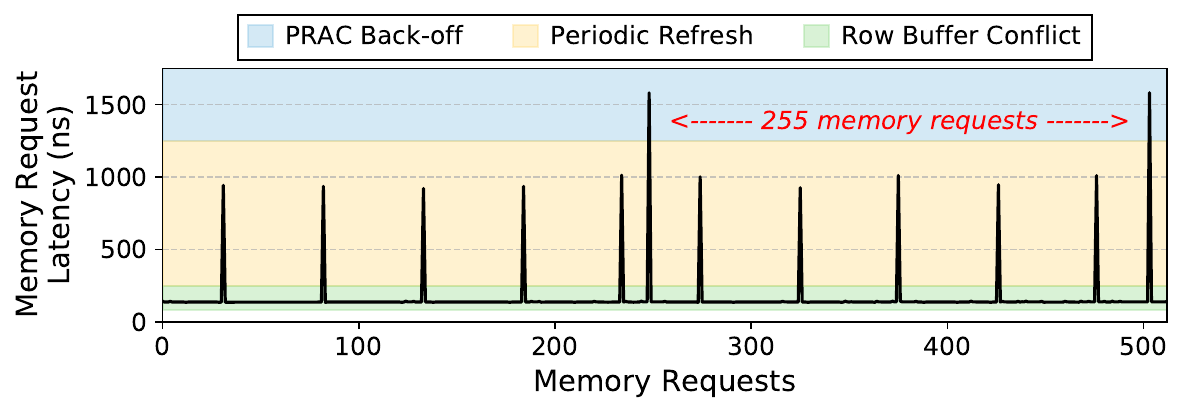}
\caption{Memory request latencies ({in ns}) of row buffer conflicts, periodic refreshes, and back-offs observed by a user space application.}
\label{fig:latency_prac}
\end{figure}

\agy{\figref{fig:latency_prac}} shows the \agy{latency measurements of \param{512} consecutive memory requests \nbcr{1}{that capture two back-off latencies} at $\aboth{} = 128$. The x-axis shows the memory requests in chronological order (older to newer from left to right), and the y-axis shows the measured latency of serving each request.}
\agy{We mark \nb{\param{three} critical latency ranges} on the y-axis based on the expected latency of a memory request under different circumstances:}
\nb{1)~row buffer conflict~\cite{keeth2001dram}, when the memory request needs to wait for the controller to issue a precharge and an activate \nb{(shown as the green range)},} 
\agy{2)~periodic refresh~\cite{liu2012raidr}, when the memory request needs to wait until a periodic refresh is completed \nb{(shown as the yellow range)},
and 
3)~PRAC back-off~\cite{jedec2024jesd795c,canpolat2024understanding}, when the memory request needs to wait until a back-off is completed \nb{(shown as the blue range)}.}

We make \param{{three}} observations from \figref{fig:latency_prac}. 
First, 
PRAC back-offs \agy{consistently cause significantly higher} latency values after accessing a DRAM row \gls{aboth} times (i.e., 255 accesses in total, as accessing one row 128 times also means the other row is accessed 127 times to create a conflict).
Second, our routine measures \nbcr{1}{an average latency of \param{1929.2 ns} for requests \nbcr{2}{that are delayed by a back-off}.} 
The observed latency is higher than the back-off latency defined in the standard (i.e., 1400~ns~\cite{jedec2024jesd795c}) \nbcr{1}{for} two reasons\nb{: (i)}~the routine measures the latency of executing one loop iteration, including the additional \nbcr{1}{instruction} latencies \nbcr{1}{within the loop}, \nb{and (ii)}~some of the memory operations are delayed by \nbcr{1}{both back-offs and periodic refreshes}, thus increasing the average latency observed. \nbcr{1}{Third, the observed back-off latency is \param{$1.9\times$} that of a memory request \nbcr{2}{that is delayed by a} periodic refresh, i.e., the next-highest latency event, on average.}\footnote{\label{footnote:ref}We \nbcr{1}{simulate} a memory controller that can postpone a periodic refresh \nbcr{1}{by a refresh interval} and issues two periodic refreshes back-to-back as it is allowed in the standard~\cite{jedec2024jesd795c}. This is observed in modern systems, aiming to improve performance by scheduling refresh at idle times~\cite{zenhammer,chang2014improving}.} 
Based on these \nbcr{1}{results}, we conclude that userspace applications can detect back-offs by comparing a measured latency \nbcr{1}{against the latency of} regular memory accesses and periodic refreshes.

\subsection{PRAC-based Covert Channel}
\label{sec:prac-attack}

\head{Attack Overview}
The sender and the receiver agree on a message encoding scheme that encodes a logic-1 bit as "back-off latency" and a logic-0 bit as "no back-off latency". 
To deterministically \nbcr{1}{induce} a back-off, the sender \agy{repeatedly activates a DRAM row, thereby increasing the activation count of the DRAM row and forcing the DRAM chip to send a back-off signal to the memory controller. To incentivize the memory controller to issue row activations as opposed to accessing data directly from the row buffer, the sender performs alternating accesses to two different rows, which causes row buffer conflicts and forces the memory controller to perform a row activation for each load request.}

\head{\agy{Utilizing The Receiver Routine to Perform Row Activations}}
The sender and \agy{the} receiver \agy{allocate} a page. \nbcr{1}{The two pages} \agy{are mapped to two different rows}
($Row_{S}$ and $Row_{R}$)
\agy{and accessing those two rows leads to row buffer conflicts.} \nbcr{1}{To ensure memory requests are served from the main memory, our} implementation \nbcr{1}{use} \texttt{clflush} \nbcr{1}{to bypasses} caches.\footnote{\label{footnote:clflush}\nbcr{1}{Alternatively, the attacker can use various other methods demonstrated by} many prior works~\cite{qiao2016new, gruss2018another, tatar2018throwhammer, lipp2018nethammer, van2018guardion} to perform RowHammer attacks by causing many activations in systems where \texttt{clflush} is a privileged instruction.}

\head{\nb{Window-Based Transmission}}
The sender and \agy{the} receiver synchronize the transmission of different bits using the wall clock. They agree on a window duration such that only one bit will be transmitted inside the window, and the next window will transmit the next bit. Thus, transmitting an N-bit message takes N transmission {windows}. 

\head{\agy{Transmitting Data Over The Covert Channel}}
During each window, the receiver accesses 
its private row ($Row_{R}$) and continuously measures the memory request latency to detect back-offs.

The sender transmits a logic-1 value by accessing its private row $Row_{S}$ until the end of the window, creating row buffer conflicts with the receiver's memory accesses. \nbcr{1}{This leads to activating} both rows and increasing their activation counts. When one of the rows' activation counter reaches \gls{aboth} activations, the receiver \ous{observes a back-off latency} and determines the transmitted bit as logic-1. To send a logic-0, the sender does not access any row, \ous{causing} the receiver to observe many row buffer hits and determine the bit is 0 (i.e., the sender is inactive). \nbcr{1}{If the receiver determines the transmitted bit before the end of the current transmission window, it sleeps until the end \nbcr{2}{of the transmission window} to avoid incrementing the activation counters.} \labelc{\label{resp:c2_2}2}\revc{\nbcr{1}{Similarly}, the sender also measures the \nbcr{1}{memory request latencies} and \nbcr{1}{sleeps} until the end of the current transmission window if it detects a back-off.}

\begin{figure}[h]
\centering
\includegraphics[width=0.95\linewidth]{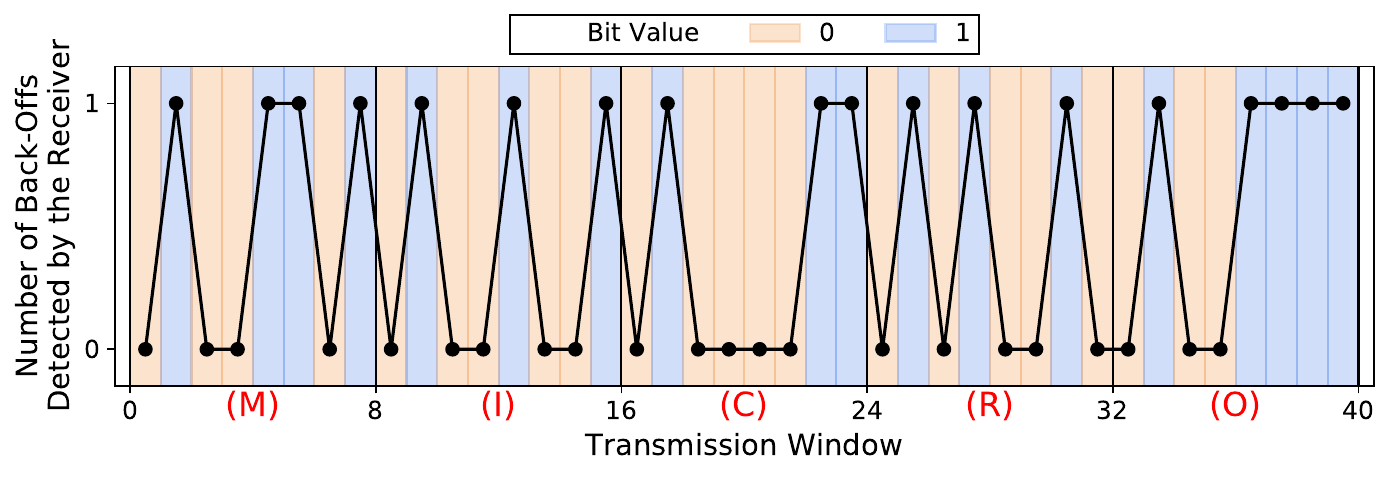}
\caption{\revb{PRAC-based covert channel demonstrating 40-bit message transmission.}}
\label{fig:prac-poc}
\end{figure}

\head{Results}\labelb{\label{resp:b1}1}
We implement a proof-of-concept covert channel attack by building a unidirectional channel (i.e., with dedicated sender and receiver processes) for \gls{aboth}$=128$. \nb{We set the window size to \param{$25~\mu s$} to account for the number of row activations needed to cause a back-off and the back-off latency.} The sender transmits the \nbcr{1}{40-bit} message \textit{"MICRO"}.   
\figref{fig:prac-poc} plots the number of accesses the receiver measures \revb{within a window} \nb{as a line plot}. \nb{The x-axis shows the transmission windows where each window is colored with the transmitted bit value, and the y-axis shows the number of back-offs detected by the receiver.}
We observe that the receiver detects a back-off during a transmission window \textit{only} when the sender is transmitting a logic-1 value. 
We conclude that the receiver successfully decodes the message after 40 transmission windows.

To evaluate the PRAC-based covert channel attack, we transmit 100-byte messages with four patterns: all 1s, all 0s, checkered 0 (i.e., 010101...01), and checkered 1 (i.e., 101010...10). 
We observe \agy{that} the attack achieves \param{39.0 Kbps} 
raw bit rate \nb{consistently} across all message patterns. \label{claim:1}

\head{\agy{Noise Analysis}}
\label{sec:c1_high}
To evaluate the impact of noise in our covert channel, we create a microbenchmark that issues memory requests \ous{targeting the DRAM bank of the covert channel} with different frequency levels and run it \ous{concurrently with the sender and the receiver processes}. Our microbenchmark increases the activation counters quickly to trigger back-off\ous{s}. We simulate different {back-off frequencies} by inserting sleep periods \nbcr{1}{\nbcr{2}{of} varying lengths} between \nb{two consecutive row activations}. We sweep the sleep duration from \param{2 $\mu s$} to \param{0.2 $\mu s$} \labelc{\label{resp:c2}2}\revc{and calculate the \nbcr{1}{noise} intensity}:

\revc{
\begin{equation}
Noise Intensity = (1 - \frac{{Sleep Duration} - MinSleep}{MaxSleep - MinSleep}) \times 99 + 1
\end{equation}}

\revc{As the sleep duration decreases, the noise intensity increases linearly.}
\nb{The lowest noise intensity level (1\%) \revc{(corresponding to 2 $\mu s$ sleep)}, represents a noise level similar to $10\times$ that of a 4-core workload consisting of highly memory-intensive SPEC2017 applications based on the back-off frequency.}

\begin{figure}[h]
\centering
\includegraphics[width=0.9\linewidth]{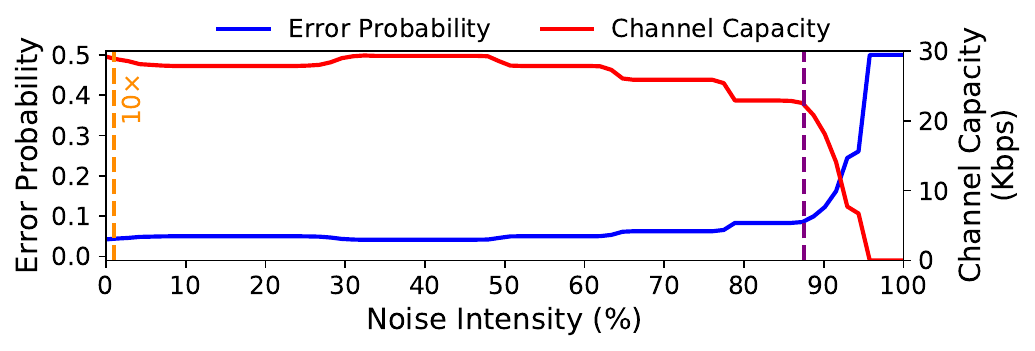}
\caption{PRAC-based covert channel's channel capacity and error probability \nbcr{1}{versus} noise intensity.}
\label{fig:prac-channel-cap}
\end{figure}

\figref{fig:prac-channel-cap} shows the error probability and the channel capacity of the covert channel \nbcr{1}{(as defined in~\secref{sec:methodology})} for different noise intensity levels. \nb{The x-axis shows the varying noise intensities as explained above. The primary (left) and secondary (right) y-axes show the error probability (plotted as the blue line), and the channel capacity (plotted as the red line), respectively.} \label{claim:3}
We make \param{three} key observations. First,  we observe \param{0.05} error probability at noise intensity \nbcr{1}{of} 1\% (shown with the orange line). At this noise level, the channel capacity is \param{28.8 Kbps}. Second, 
the covert channel's capacity remains high (>20.7Kbps) \nbcr{1}{since} error probability remains below 0.1 until a very high noise intensity of 88\%. 
Third, as noise intensity increases above $88\%$, the error probability increases, degrading channel capacity. 
This noise level is similar to having an aggressive memory performance attack~\cite{moscibroda2007memory}, \nb{which is not desired in memory systems already}. 
\agy{Based on these observations, we} conclude that the PRAC-based covert channel attack maintains its \nbcr{1}{channel} capacity until very high noise intensity values. 

\begin{figure}[h]
\centering
\includegraphics[width=0.9\linewidth]{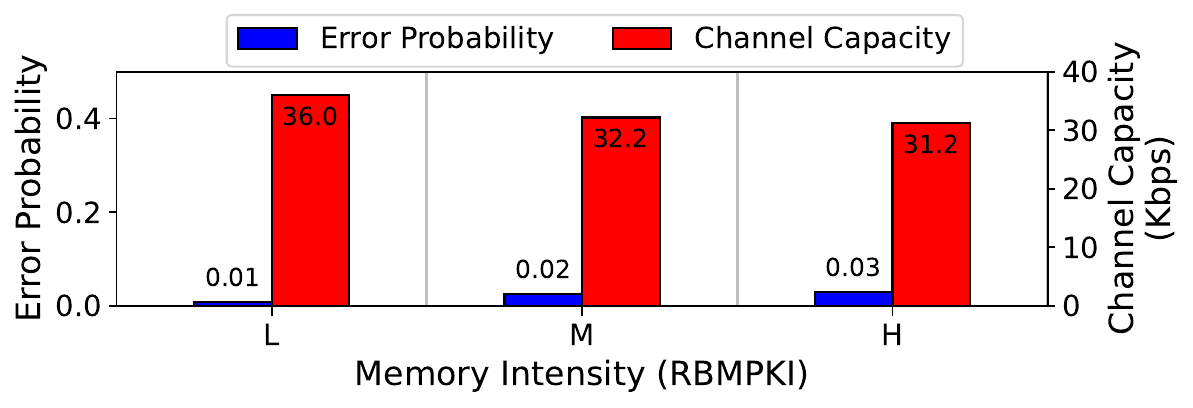}
\caption{\revcommon{PRAC-based covert channel's capacity and error probability with concurrently running SPEC2017 \nbcr{1}{applications.}}}
\label{fig:prac-spec}
\end{figure}

\head{\revcommon{Application-Induced Noise}}\labelcq{\label{resp:cq3_1}3\\C, E}
\revcommon{We evaluate the impact of interference from concurrently running applications on error probability and channel capacity of \nbcr{1}{our} PRAC-based covert channel attack. We run the sender and receiver processes concurrently with SPEC2017~\cite{spec2017} applications that exhibit low (L), medium (M), and high (H) memory intensity. We measure the memory intensity with the row buffer misses per kilo instructions (RBMPKI) metric and categorize benchmarks into three categories. \figref{fig:prac-spec} shows error probability (primary y-axis) and channel capacity (secondary y-axis) for increasing memory intensities. We make two key observations. First, as memory intensity increases, error probability increases slightly due to row buffer conflicts caused by concurrently running applications. Due to \nbcr{1}{application-induced} interference, the receiver can activate its row even when the sender is inactive and gradually increase the activation counter, which can result in additional back-offs. Second, at the highest memory intensity level, the PRAC-based covert channel's capacity is 31.2 Kbps due to the 0.03 error probability. We conclude that real-world application-induced interference does not prevent an attacker from exploiting \X{} and only reduces channel capacity.}

\head{Multibit Covert Channels}
\label{sec:multibit}
\agy{We extend the PRAC-based covert channel to \nbcr{1}{support ternary and quaternary transmissions (i.e., 1.58 and 2 bits per transmission, respectively, compared to 1-bit binary transmissions).}
To achieve this, we vary the sender process' memory intensity, such that the receiver observes a back-off latency after performing a \nbcr{1}{specific} number of memory accesses.} \nb{This way, the message can be encoded using the number of accesses that the receiver performs until a back-off (or until the end of the transmission window if no back-off \nbcr{1}{occurs}).} \agy{Our evaluation of transmitting 32-byte messages shows that doing so achieves \nb{raw bit rates of} \param{39.0}, \param{61.7}, and \param{76.8} Kbps for binary, ternary, and quaternary configurations, respectively. This \nbcr{1}{higher} \nb{raw bit rate} comes at the cost of reduced \nbcr{1}{noise} tolerance, such that ternary and quaternary configurations} \nb{\nbcr{1}{exhibit} higher error \nbcr{1}{probabilities} of \param{0.04} and \param{0.29}, \nbcr{1}{resulting in channel capacities of} \param{46.7 Kbps} and \param{10.1 Kbps}, respectively. We conclude that \X{} covert channels support multibit transmissions with the tradeoff of being more susceptible to noise. }

%% file: sections/05_1_casestudy.tex
\section{Case Study 2: RFM-based Covert Channel}
\label{sec:case_two}
\subsection{Periodic RFM Overview}
As a second case study, we construct a second covert channel exploiting a memory controller-based {defense} technique called \gls{prfm}, as described in early DDR5 standards~\cite{jedec2020jesd795}. \gls{prfm} uses a tracking mechanism in the memory controller to count the number of activations for each DRAM bank. When a DRAM bank's counter reaches a predefined threshold value called \gls{rfmth}, the memory controller issues an \gls{rfm} command. In our implementation, we assume \gls{rfmth} is 40, which is one of the \gls{rfmth} values supported in the standard~\cite{jedec2024jesd795c}.

\subsection{RFM-induced Memory Latencies}
\label{sec:rfm_characterization}

To observe the RFM-induced latencies from userspace applications, we use the routine presented in Listing~\ref{ls:latency_measure} to 1)~trigger \gls{prfm} to issue \gls{rfm} commands and 2)~measure the memory request latencies within the loop.   
We make \param{two} observations from this experiment. First, \gls{rfm} commands consistently \nb{cause significantly higher} latency values after accessing rows within the same bank \gls{rfmth} times, i.e., $41.8$ accesses on average.
Second, the routine measures the average latency of a memory request coinciding with an \gls{rfm} command as \param{{419.1 ns}} in our testing environment, which is higher than the \gls{rfm} latency defined by the standard (e.g., 295 ns~\cite{jedec2024jesd795c}) due to 1) measuring the execution time of the whole loop and 2) having a portion of the \gls{rfm} commands coinciding with periodic refreshes. 
We conclude that a userspace application can detect \gls{rfm} commands by comparing the measured latency to the latency of regular memory requests.

\subsection{RFM-based Covert Channel}
\label{sec:rfm_attack}
\head{Attack Overview}
\nbcr{1}{The sender and the receiver leverage \gls{prfm}-induced high latencies to transmit messages. \gls{prfm}'s activation counters are noisy because all accesses to the same bank increment the same activation counter. Therefore, attack processes (and other concurrently-running processes) can trigger preventive actions unintentionally (i.e., by accessing a bank \gls{rfmth} times) during the attack. To account for the noise from additional preventive actions, the sender transmits each bit multiple times and the receiver determines the bit value by counting the number of \gls{rfm}s within each transmission window.}

The sender and the receiver each allocate one page in separate DRAM rows ($Row_{S}$ and $Row_{R}$) in the same DRAM bank.\nbcr{1}{To ensure memory requests are served from the main memory, our implementation leverages \texttt{clflush} \nbcr{1}{and bypasses} caches.}\footref{footnote:clflush} 
The sender and receiver synchronize between the transmission of different bits using the wall clock (as described in~\secref{sec:prac-attack}). Transmitting an N-bit message takes N transmission windows. 

\head{Transmitting Data Over The Covert Channel} In each window, the receiver accesses $Row_{R}$ and continuously measures the memory request latencies to count the number of \gls{rfm} commands. As the bank activation counters are noisy, counting the number of \gls{rfm}s per transmission window increases the robustness of the attack against \nbcr{1}{interference.}

The sender sends one bit by \nbcr{1}{either} increasing the bank activation counter (logic-1) or not (logic-0). To transmit a logic-1 value, the sender accesses $Row_{S}$ to create row buffer conflicts with the receiver and increase the bank activation counter until the window ends. \nbcr{2}{This causes \gls{prfm} to issue \gls{rfm} commands, which the receiver observes as multiple preventive actions within the transmission window.} To send a 0, the sender sleeps until the end of the transmission window, reducing the number of row buffer conflicts with the receiver. This way, the receiver measures fewer \gls{rfm} commands \nbcr{1}{due to fewer activations in the same bank}. At the end of a window, the receiver compares the number of \gls{rfm}s to a predetermined threshold ($T_{recv}$) to determine the bit value.

\head{Results}
We implement a proof-of-concept of the covert channel attack by building a unidirectional channel (i.e., with dedicated sender and receiver processes) and setting $T_{recv}=3$ for \gls{rfmth}$=40$. \nb{We set the window duration to $20~\mu s$ to account for the number of accesses needed to trigger \gls{rfm} multiple times and the \gls{rfm} latency.} The sender transmits the 40-bit message \textit{"MICRO"}.

\begin{figure}[h]
\centering
\includegraphics[width=\linewidth]{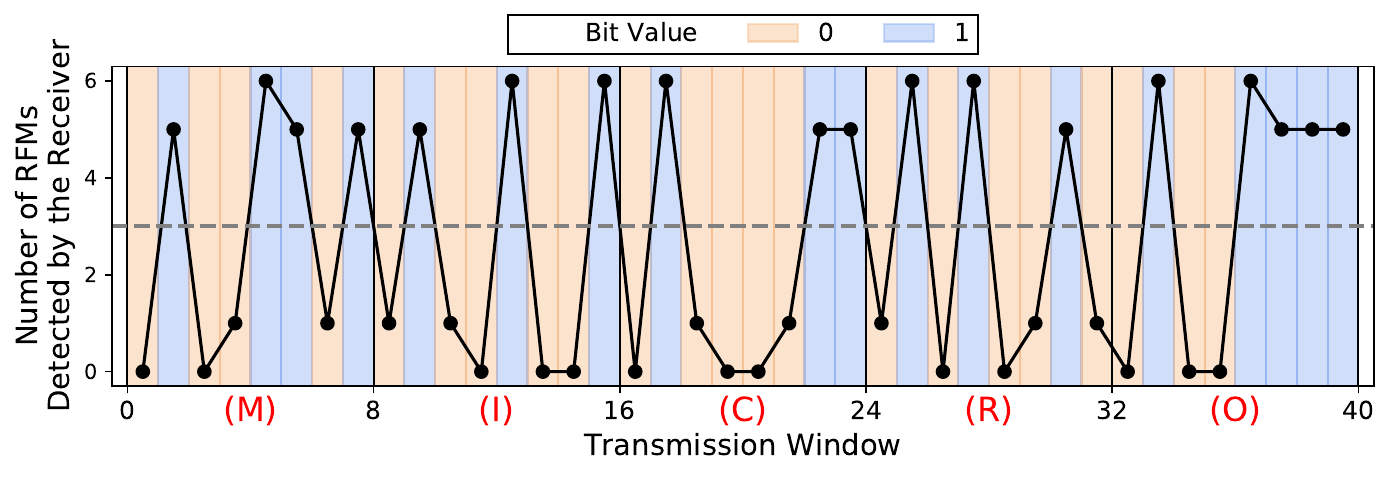}
\caption{RFM-based \X{} Covert Channel demonstrating 40-bit message transmission.}
\label{fig:rfm-poc}
\end{figure}

\figref{fig:rfm-poc} plots the number of \gls{rfm} commands the receiver detects for each transmission window \nb{as a line plot}. \nb{The x-axis shows the transmission windows where each window is colored with the transmitted bit value, and the y-axis shows the number of \gls{rfm}s the receiver detects within a window.} We observe that the receiver successfully decodes the message after 40 transmission windows.

To evaluate the proof-of-concept \gls{rfm}-based covert channel attack, we transmit 100-byte messages with four patterns: all 1s, all 0s, checkered 0 (i.e., 01...01), and checkered 1 (i.e., 10...10). We observe that the attack achieves \param{48.7 Kbps} raw bit rate on average across all message patterns. \label{claim:2}

\begin{figure}[h]
\centering
\includegraphics[width=0.95\linewidth]{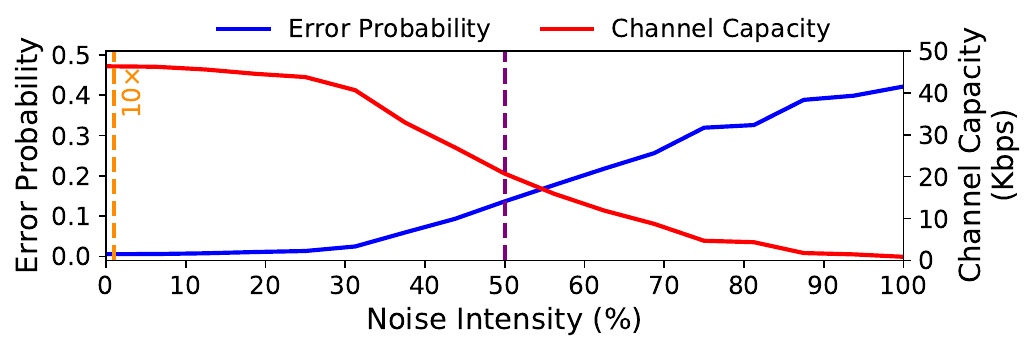}
\caption{RFM-based covert channel's channel capacity and error probability with increasing noise intensities.}
\label{fig:rfm-channel-cap}
\end{figure}

\head{Noise Analysis} \label{sec:c1_high2} To evaluate the impact of noise in our covert channel, we run a microbenchmark that issues memory requests with different frequency levels concurrently with sender and receiver processes (as described in \secref{sec:prac-attack}). Our microbenchmark induces row activations to increase activation counters quickly and trigger RFM. \figref{fig:rfm-channel-cap} shows error probability and channel capacity \nbcr{1}{(as described in~\secref{sec:methodology})} for different noise intensity levels. \nb{The x-axis shows \nbcr{2}{noise intensity}. The primary (left) and secondary (right) y-axes show error probability (plotted as the blue line) and channel capacity (plotted as the red line), respectively.}
\nb{The lowest noise intensity level (1\%), represents a noise level similar to $10\times$ that of a 4-core workload consisting of highly memory-intensive SPEC2017 applications \nbcr{1}{in terms of \gls{rfm} \nbcr{2}{frequency}}.}

We make \param{three} key observations. First, we observe a \param{<0.01} error probability at the lowest noise intensity level (indicated by the orange line). At this noise level, channel capacity is \param{46.3 Kbps}. Second, channel capacity remains high (>20.7 Kbps) as error probability remains below 0.1 until a noise intensity of 50\% (indicated by the purple line).  Third, as noise intensity increases above 50\%, channel capacity reduces rapidly due to the noise generator microbenchmark consistently triggering many \gls{rfm}s within each transmission window. 
We conclude that \nbcr{1}{our} RFM-based covert channel maintains its capacity until high noise intensity values. \label{claim:4}

\begin{figure}[h]
\centering
\includegraphics[width=0.9\linewidth]{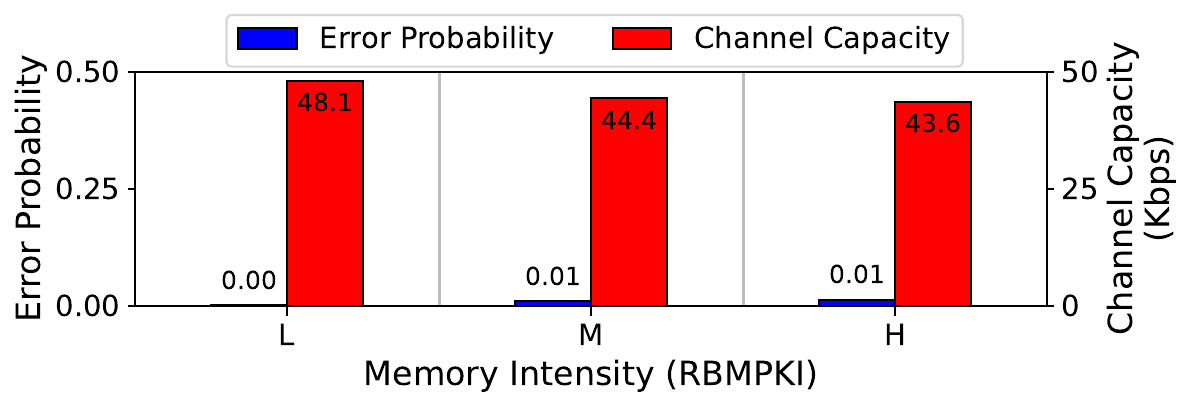}
\caption{\revcommon{RFM-based covert channel's capacity and error probability with concurrently running SPEC2017 \nbcr{1}{applications.}}}
\label{fig:rfm-spec}
\end{figure}

\head{\revcommon{Application-Induced Noise}}\labelcq{\label{resp:cq3_2}3\\C, E}
\revcommon{\figref{fig:rfm-spec} shows error probability and channel capacity for increasing memory intensities of concurrently running SPEC2017~\cite{spec2017} applications (using a similar style as~\figref{fig:prac-spec}). We make two key observations. First, as memory intensity increases, error probability increases slightly due to the row buffer conflicts caused by the concurrently running application. Second, at the highest memory intensity level, RFM-based covert channel's capacity is 43.6 Kbps due to \nbcr{1}{an error probability of 0.01}. We conclude that real-world application-induced interference does not prevent LeakyHammer and only impacts channel capacity slightly.}

\head{Comparing PRAC-Based and RFM-Based Covert Channels}
We \agy{identify}
\param{two} differences between \agy{PRAC- and RFM-based} covert channels.
First, the \gls{rfm}-based covert channel has a higher raw bit rate \nbcr{1}{(e.g., 48.7 Kbps)}.
\agy{This is because an RFM command can be triggered more frequently (e.g., once every 32 to 80 activations instead of $\ge$128) 
and has smaller latency (e.g.,~\SI{295}{\nano\second} instead of \SI{1.4}{\micro\second}) compared to back-off~\cite{jedec2024jesd795c}.}
\nb{Second, the \gls{prac}-based covert channel is more robust to noise than the \gls{rfm}-based covert channel. 
\nbcr{1}{This is because \gls{prfm} 1)~uses bank-level activation counters that aggregate all row activations \nbcr{2}{to} the same bank and 2)~requires fewer activations to trigger a preventive action. \nbcr{2}{Concurrently-running} attack and non-attack processes can unintentionally cause preventive actions with fewer activations in the same bank and lead to \nbcr{2}{errors in transmissions}.} 
We conclude that the \gls{prac}-based covert channel provides higher channel capacity in noisy environments than the \gls{rfm}-based covert channel, due to its robustness to noise.} 

%% file: sections/05_2_casestudy.tex
\section{Case Study 3: PRAC-based Side Channel}
\label{sec:sidechannel}
We build and evaluate a website fingerprinting attack~\cite{hintz2002fingerprinting,sun2002statistical} \nbcr{1}{where} an attacker detects which website a victim user visits in a system \nbcr{1}{employing} PRAC.  

\head{Website Fingerprinting Attack Model}
We assume the \nbcr{1}{victim} user is accessing a sensitive website using a web browser. The attacker executes on the same system and can \textit{only} observe the latencies of its own memory requests. It has access to 1)~fine-grained timers, such as the \texttt{rdtsc} instruction~\cite{intelmanual2016}, and 2)~instructions to flush cache blocks, such as the \texttt{clflush} and \texttt{clflushopt} instructions~\cite{intelmanual2016}. We use timer \nbcr{1}{(using \texttt{m5\_rpns()} in our simulation environment)} and cache-flush instructions. The attacker can also use fine-grained timers designed for web browsers~\cite{schwarz2017fantastic,gras2017aslr} and eviction sets to flush cache blocks~\cite{gruss2016rowhammer} where such instructions are unavailable.

The attacker allocates DRAM rows \ous{and measures} memory request latencies \ous{to} detect \ous{back-offs} as described in \secref{sec:observe_prac}. The attacker does \emph{not} need to colocate \nbcr{1}{its data with the victim process's data at the row level} \ous{because} back-offs are \ous{performed at \nbcr{1}{channel} granularity} (i.e., all memory requests targeting the same \nbcr{1}{channel} observe the latency increase). The attacker can either colocate \nbcr{1}{its data} \nbcr{1}{with the victim} in one \nbcr{1}{channel} or allocate rows across all \nbcr{1}{channels} after \nbcr{1}{partially} reverse engineering the DRAM address mapping \nbcr{1}{with existing reverse engineering methods~\cite{pessl2016drama,zenhammer,wang2020dramdig,helm2020reliable,heckel2023reverse}.} %

\head{Fingerprinting Routine}
To collect accurate fingerprints of websites, the fingerprinting routine should avoid triggering back-offs to observe the website's back-off behavior. We construct a routine \nbcr{1}{that \nbcr{2}{avoids} triggering preventive actions as} described in Listing~\ref{ls:fingerprint}. \nbcr{2}{The routine} allocates N \textit{test rows}. In each iteration, the routine accesses a test row T times (lines 10-16), where T is smaller than \gls{aboth}, and measures the latency of the memory requests (line 14). The routine accesses another row in the next iteration.
To minimize interference \nbcr{1}{(i.e., another application incrementing the same activation counter as the attacker and leading to preventive actions)}, the routine can allocate each test row fully (i.e., no other application allocates a page within the same row) or reduce T.
We empirically set N and T to cover the execution time of \ous{loading the} website.
{The routine collects timestamps for each measurement by calculating the elapsed time since the start \nbcr{1}{of the attack}.}

\begin{figure}[h]
\begin{lstlisting}[style=customc, label = ls:fingerprint, caption = Fingerprinting routine collecting memory request latency measurements.]
 // row_ptrs array has N address pointers
 // that point to N different DRAM rows
 // T = PRAC back-off threshold - 1
 vector<uint64_t> measure (vector<char*>& row_ptrs) {
    vector<uint64_t> latency(ITERATIONS*T,0);
    // get start timestamp
    uint64_t start = m5_rpns();
    for (int i = 0; i < ITERATIONS; i++) {
        int a = i % row_ptrs.size();
        auto row_ptr = row_ptrs[a];
        // access the same target row T times 
        for (int j = 0; j < T; j++) {
            // flush the target cacheline from caches
            clflush(row_ptr); 
            // access the target row
            *(volatile char*)row_ptr;
            // get end timestamp
            uint64_t end  = m5_rpns();
            // record the measured latency
            latency[i] = end-start;
            start = end;
        }
    }
    return latency;
}
\end{lstlisting}
\end{figure}

\head{Data Collection}
To create fingerprints for each website, we collect memory request latency traces when a web browser loads the website by running the fingerprinting routine concurrently with the web browser. 
The resulting trace is similar to a memorygram~\cite{oren2015spy}, a trace of cache access latencies measured by cache-based fingerprinting attacks~\cite{shusterman2019robust,oren2015spy}.
We generate traces for different websites using a memory trace generator tool based on Intel Pin~\cite{intelpin}. During trace generation, we load \nbcr{1}{each} website using the same web browser and keep it open for 20 seconds before closing the window. We collect 50 traces per website by repeating this procedure. We use the collected memory traces to simulate the web browser in our simulation environment with PRAC and run it concurrently with the fingerprinting routine described above. We fingerprint \labelcq{\label{resp:cq2}2\\B, E}\revcommon{40 top visited websites}\footnote{\revcommon{We evaluate the following websites: aliexpress, amazon, apple, baidu, bilibili, bing, canva, chatgpt, discord, duckduckgo, facebook, fandom, github, globo, imdb, instagram, linkedin, live, naver, netflix, nytimes, office, pinterest, quora, reddit, roblox, samsung, spotify, telegram, temu, tiktok, twitch, weather, whatsapp, wikipedia, x, yahoo, yandex, youtube, zoom}.} similar to prior works~\cite{shusterman2019robust, rimmer2017automated}.

\begin{figure}[h]
\centering
\includegraphics[width=0.9\linewidth]{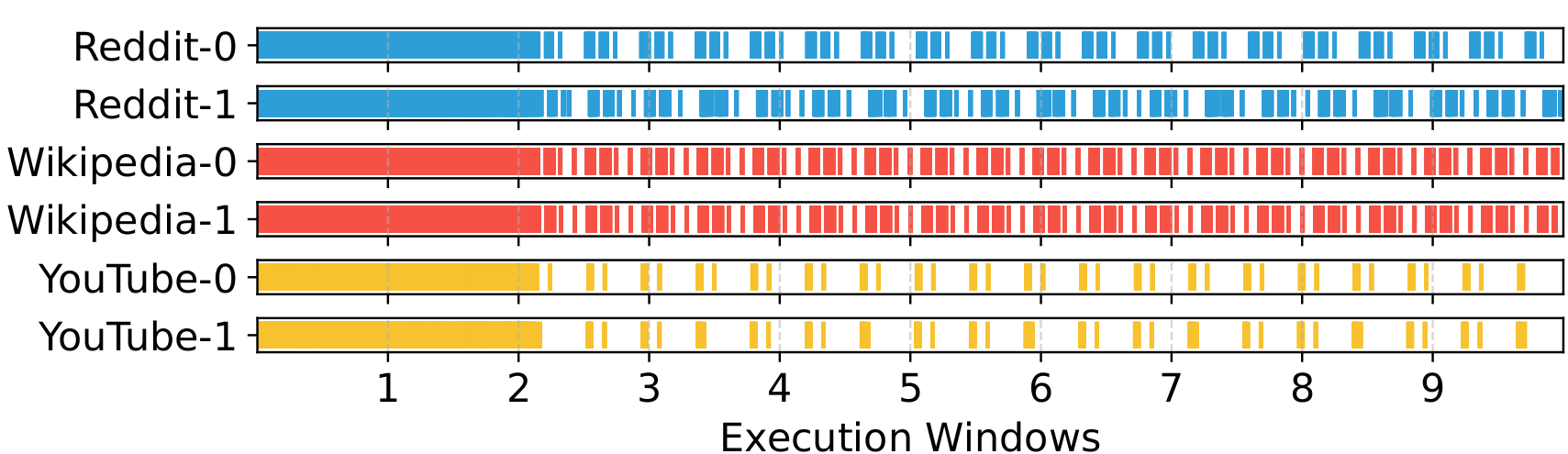}
\caption{Examples of website fingerprints based on back-offs. Each fingerprint displays back-offs as colored strips.}
\label{fig:rfmgram}
\end{figure}

\sloppy
\figref{fig:rfmgram} shows two sample fingerprints for three different websites: Wikipedia, Reddit, and YouTube. Each fingerprint displays \ous{back-offs} as colored strips across execution time (x-axis). We split the execution time into a fixed number of windows that we call \emph{execution windows} (y-axis). We make \param{three} observations.
First, different fingerprints of one website tend to have \ous{similar number and frequency of back-offs} (e.g., YouTube fingerprints).
Second, fingerprints from different websites tend to have different characteristics across execution windows.
For example, Reddit and YouTube fingerprints show different frequencies of \ous{back-offs}, starting from the execution window 2.
Third, fingerprints from different websites share similarities for smaller portions of the execution time (e.g., execution window 0 and 1) \nb{potentially} due to the browser's memory accesses executed regardless of the loaded website.
However, these similarities do not dominate the whole execution time. 
Based on these observations, we conclude that a website's back-off trace is a good candidate for identifying websites in a fingerprinting attack.

\head{Attack Overview}
\ous{The attacker determines the website a victim visits in three steps.
First, before the attack, the attacker creates a back-off trace database using the fingerprinting routine on} various websites.
\ous{Second, during the attack, the attacker} uses the fingerprinting routine to get the trace of the {loaded website}.
\ous{Third, the attacker} analyzes the \ous{collected website} fingerprint using \ous{the back-off trace} database to determine which website was loaded.

Our proof-of-concept implementation uses a machine-learning approach to classify websites. We create \nbcr{1}{training} and test datasets and train classical machine learning models \nbcr{1}{(decision tree~\cite{breiman1984classification}, random forest~\cite{breiman2001random}, gradient boosting~\cite{friedman2001greedy}, k-nearest neighbors~\cite{cover1967nearest}, support vector machines~\cite{cortes1995support}, logistic regression~\cite{cox1958regression}, Ada-boost~\cite{freund1997decision}, and perceptron~\cite{rosenblatt1958perceptron})} \nbcr{1}{on} labeled samples.
For the training, we collect \ous{features from} {each consecutive back-off pair: (i)~time between \nbcr{1}{two signals in the pair}, (ii)~the time between the start of the current pair and the end of the previous pair, and (iii)~the average of the timestamps within the pair.}
\nbcr{1}{During inference, our implementation applies the trained classifiers to} unseen and unlabeled \nbcr{1}{back-off traces} and outputs \nbcr{1}{the predicted website} label. 

\begin{figure}[h]
\centering
\includegraphics[width=0.85\linewidth]{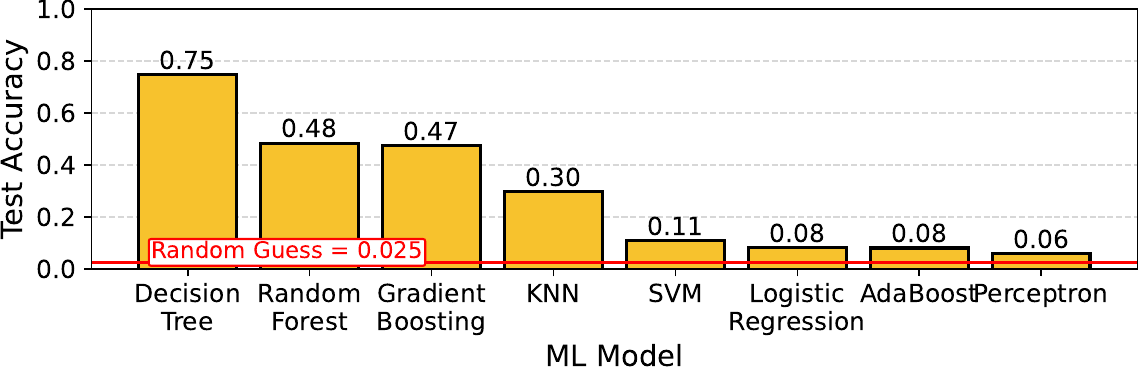}
\caption{\revcommon{Accuracy of classifiers guessing websites.}}
\label{fig:accuracy}
\end{figure}

\head{Results} We evaluate the website fingerprinting attack assuming \gls{nrh}$=64$. \figref{fig:accuracy} shows the accuracy of classifiers as the rate of their correctly assigned labels. The red line represents the random guess chance, which is \revcommon{{0.025} since the dataset has 40 websites}.

\revb{Table~\ref{tab:dt-macro-results} shows the F1, precision, and recall scores~\cite{chinchor1992muc,rijsbergen1979information} of\labelb{\label{resp:b2_2}2} the best-performing model (i.e., decision tree~\cite{breiman1984classification}), employing a 10-fold cross-validation method~\cite{kohavi1995study}. For each fold, we calculate F1-score, precision, and recall, and report the average and standard deviation.}

\begin{table}[h]
\centering
\caption{\revb{Website fingerprinting performance employing 10-fold cross-validation: F1 (\%), Precision (\%), and Recall (\%).}}
\label{tab:dt-macro-results}
\begin{tabular}{|l||c|c|c|} 
\hline
\textbf{Model} & \textbf{F1}     & \textbf{Precision} & \textbf{Recall}  \\
\textbf{Name}  & Mean ($\sigma$) & Mean ($\sigma$)    & Mean ($\sigma$)  \\ 
\hhline{|=::===|}
Decision Tree  & 71.8 (4.2)      & 74.1 (4.4)         & 72.4 (4.2)       \\
\hline
\end{tabular}
\end{table}

We observe that classical machine learning models classify websites with up to \revcommon{{{75\%} accuracy ({$30\times$}} the accuracy of a random guess)} using a back-off trace. 
\ous{We conclude that 
our proof-of-concept demonstrates it is possible to guess which website the victim loads by analyzing back-off traces using relatively simple machine learning models and features.
We leave the exploration of stronger machine learning models (e.g., transformers~\cite{vaswani2017attention}) for future work.}

\head{\revcommon{Application-Induced Noise}}\labelcq{\label{resp:cq3_3}3\\C, E}
\revcommon{We evaluate the impact of application-induced noise on the website fingerprinting attack \nbcr{1}{by running} SPEC2017 benchmarks~\cite{spec2017} \nbcr{1}{concurrently} with the attacker and browser processes. We observe that with application-induced noise, the attack achieves a 66.1\% classification accuracy. We conclude that concurrently running applications impact classification accuracy but do not prevent the attack.}

%% file: sections/12_comparison_new.tex
\section{{\X{} \nbcr{3}{versus} Prior Work}}
\label{sec:comparison_against_others}

LeakyHammer showcases a new timing channel that exploits RowHammer defenses. LeakyHammer has \param{two} features that differentiate it from the existing \nbcr{1}{cache-} and DRAM-based attacks.

\head{{Attack Scope}}
LeakyHammer's attack scope is wider than that of existing cache- and DRAM-based timing channels.
Existing high-throughput cache- and DRAM-based timing channels~\cite{maurice2017hello,yarom2014flush+,gruss2016flush+,liu2015last,xiong2020leaking,lipp2020take,pessl2016drama} require the attacker and the victim (or the sender and the receiver) to \nb{either} 
1) share memory pages (e.g., Flush+Flush~\cite{gruss2016flush+},  Flush+Reload\cite{yarom2014flush+}), 2) share the same CPU (e.g., cache-based attacks~\cite{maurice2017hello,yarom2014flush+,gruss2016flush+,liu2015last,xiong2020leaking,lipp2020take}), or 3) colocate \nbcr{1}{data} in the same DRAM bank (e.g., row buffer conflict-based attacks~\cite{pessl2016drama,dramaqueen}). LeakyHammer can observe latency differences either at \nbcr{1}{channel} \nbcr{2}{granularity} (e.g., PRAC back-off~\cite{jedec2024jesd795c}) or at bank group \nbcr{2}{granularity} (e.g., RFM~\cite{jedec2020jesd795}). This effectively expands the scope of the attack by both enabling cross-CPU attacks and relaxing the colocation requirements. 
\reva{This enables attackers to exploit \X{} in systems that employ bank partitioning~\cite{muralidhara2011reducing,liu2014bpm,volos2024principled,suzuki2013coordinated,yun2014palloc,liu2012software,jeong2012balancing,xie2014improving}, in contrast to same-bank attacks (e.g., row buffer-based attacks~\cite{dramaqueen,pessl2016drama}). Systems that employ bank partitioning 1) reduce interference at the bank level across different applications and improve fairness, which is shown to be beneficial in multicore settings by prior works~\cite{yun2014palloc,liu2014bpm,liu2012software,jeong2012balancing,xie2014improving,muralidhara2011reducing}, and 2) ensure security against same-bank attacks~\cite{volos2024principled,suzuki2013coordinated}.}

\head{{\atb{Effectiveness} of Existing Mitigations}}
{Existing defenses against row buffer-based covert channels are ineffective against LeakyHammer. A very simple and effective defense primitive against DRAMA~\cite{pessl2016drama} is to have a strictly closed-row policy where a row is immediately precharged after every access. This solution does \textit{not} \atb{mitigate} LeakyHammer. A similar constant-time enforcement solution \nbcr{1}{would require} all memory requests to \nbcr{1}{exhibit a memory latency including} the preventive action latency (e.g., 295 ns for RFM and 1400 ns for PRAC back-off) on top of the maximum memory latency. \nbcr{1}{This approach would} incur significant performance overheads \nbcr{1}{due to high latency of \rh{} preventive actions}.}

\subsection{Comparison Against {\nb{Existing} DRAM-Based \nb{Timing Channels}}} \labelcq{\label{resp:cq1}{1\\A, B, E}}
\revcommon{We compare \X{} against the state-of-the-art DRAM-based timing channel, DRAMA~\cite{pessl2016drama}, \nbcr{1}{which} leverages the DRAM row buffer as a timing channel by detecting row buffer hits and conflicts.}

\head{Information Leakage Model}
\revcommon{To establish a covert communication channel, DRAMA colocates the sender and receiver in the same DRAM bank and transmits messages by creating conflicting memory accesses and measuring the memory access latencies (i.e., uses row buffer conflict-based timing differences). Even though \X{} uses a similar access pattern, the goal of the sender and receiver is to increase activation counters quickly and transmit messages with \rh{} defense-induced higher latencies (e.g., 10$\times$ of a row buffer conflicting memory access in our evaluated system for PRAC~\cite{jedec2024jesd795c}). \nbcr{1}{In contrast to DRAMA, \X{} does not require the sender and the receiver \nbcr{2}{to share} the same bank.} The sender {can simply alternate between two rows within one bank (e.g., bank~0) to increase the activation counters}. The receiver \nbcr{1}{can measure the latency of accesses to a row in another bank (e.g., bank~1) to detect the \rh{} preventive actions.}}

\revcommon{A set of DRAMA side channel attacks~\cite{pessl2016drama} colocate \nbcr{1}{their data} with \nbcr{1}{the victim process's data} \nbcr{2}{at} DRAM row granularity (i.e., share the same DRAM row \nbcr{1}{with the victim}) and leak whether (or when) the victim activates that specific DRAM row.
In contrast, \X{} colocates \nbcr{1}{data} with the victim \nbcr{1}{at} \nbcr{1}{channel} or bank group \nbcr{2}{granularity} and leaks the information that the victim accesses DRAM with an access pattern that triggers the RowHammer defense.}

\head{\revcommon{Attack Capabilities}}
\revcommon{We describe the attack capabilities of \X{}-PRAC, \X{}-RFM, and DRAMA~\cite{pessl2016drama} (in terms of the leaked information) assuming different colocation \nbcr{2}{granularitie}s in Table~\ref{tab:attack_capability}.}

\begin{table*}[ht!]
\centering
\caption{Information leaked by \X{} and DRAMA~\cite{pessl2016drama}}
\resizebox{\textwidth}{!}{
\begin{tblr}{
  vline{2} = {-}{},
  hline{1-2,5} = {-}{},
}
                                            {Colocation \nbcr{2}{Granularity}}
                                            & \textbf{\nbcr{1}{Channel} / Bank Group}                                            & \textbf{Bank}                                                         & \textbf{Row}                                                                 \\
LeakyHammer-PRAC                            & {Victim triggered a preventive action\\(i.e., memory access pattern)} & {Victim triggered a preventive action\\(i.e., memory access pattern)} & {Number of times the victim\\has activated the row (i.e., activation count)} \\
LeakyHammer-RFM                             & {Victim triggered a preventive action\\(i.e., memory access pattern)} & {Number of times the victim\\has activated a row in the same bank}    & {Number of times the victim\\has activated a row in the same bank}           \\
DRAMA~\cite{pessl2016drama} & N/A                                                                   & {Victim accessed a conflicting row\\or the same row}      & {Victim accessed a conflicting row\\or the same row}                      
\end{tblr}
}
\label{tab:attack_capability}
\end{table*}

\revcommon{We make three key observations. First, at \nbcr{1}{channel} or bank group \nbcr{2}{granularity}, only \X{} can leak information. \X{} leaks \nbcr{1}{if} the victim \nbcr{1}{has} a specific access pattern (e.g., \nbcr{1}{accessing} the same DRAM row \gls{aboth} times for \gls{prac} or the same DRAM bank \gls{rfmth} times \gls{rfm}), as demonstrated in~\secref{sec:sidechannel}. 
Second, if the attacker colocates \nbcr{1}{its data} with the victim at the same \nbcr{2}{granularity} as the activation counter granularity (i.e., \nbcr{1}{sharing the same row for} \gls{prac}, and the \nbcr{1}{same} bank for \gls{rfm}), the attacker can leak how many times the victim increments a specific counter (i.e., activation counter value). Thus, \X{}-PRAC can leak how many times the victim activates a specific row by \nbcr{1}{sharing a row with the victim}, and \X{}-RFM can leak how many times the victim activates \nbcr{1}{rows} in a DRAM bank by \nbcr{1}{sharing a bank with the victim}. \nbcr{1}{By leaking the activation counter values}, the attacker can leak multiple bits at once \nbcr{1}{(e.g., log(\gls{aboth}) or log(\gls{rfmth}))}.
Third, at bank \nbcr{2}{granularity} and row \nbcr{2}{granularity}, DRAMA leaks whether the victim is accessing a different row in the same DRAM bank (i.e., 1 bit of information).}

We demonstrate an attack leaking an activation counter value in a system with \gls{prac} in our simulation-based evaluation environment (\secref{sec:methodology}). At \gls{aboth}$=$128, we observe that an attacker can leak a counter value of 7 bits in $13.6~\mu s$, on average, achieving 501 Kbps leakage throughput. We provide much more detail on this attack in the extended version of this paper~\cite{bostanci2025understanding}.

%% file: sections/10_sensitivitystudy.tex
\section{{Sensitivity Studies}}
\label{sec:sensitivity}
In this section, we analyze 1) the impact of preventive action latency on \X{} covert channels (\secref{sec:sweeping_rfm}, \secref{sec:reduced_latency}) and 2) the impact of a larger cache hierarchy and data prefetching on \X{} attacks (\ref{sec:cache}).

\subsection{{\nbcr{1}{Effect of} the Number of RFM Commands\\during Back-off Period}}
\label{sec:sweeping_rfm}

\begin{figure}[!ht]
    \centering
    \includegraphics[width=0.9\linewidth]{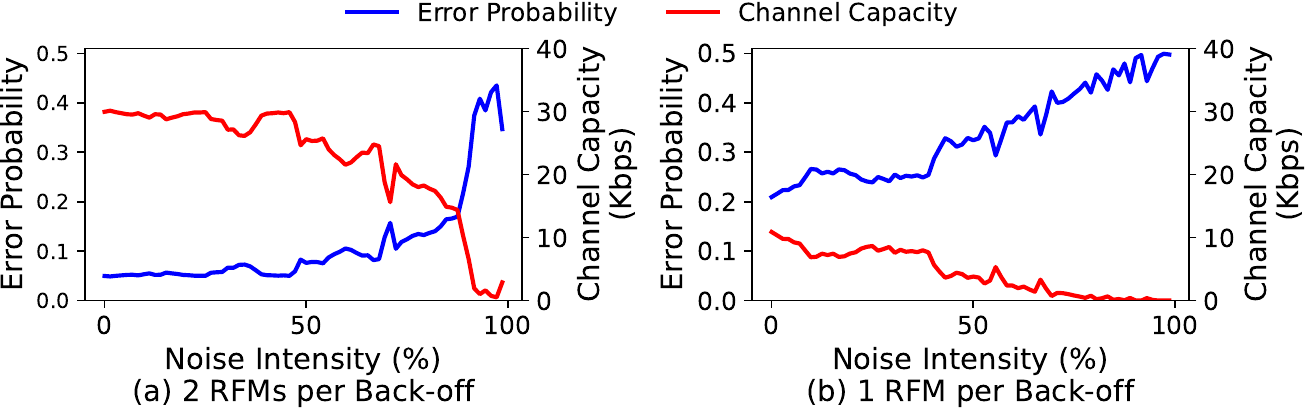}
    \caption{Error probability and channel capacity of LeakyHammer with a) 2 RFMs and b) 1 RFM per back-off.}
    \label{fig:rfm_comparison}
\end{figure}

\nbcr{1}{We evaluate the performance of LeakyHammer on a system with PRAC, where we implement back-offs using either 1)~two RFMs or 2)~one RFM.} For simplicity, we assume \nbcr{1}{periodic refreshes are not postponed} in this section.\footnote{Allowing postponing periodic refreshes once results in similar trends: with 2-RFM back-offs, the back-off latency overlaps with the periodic refresh, whereas with 1-RFM back-offs, there is a larger gap between the two.}
\figref{fig:rfm_comparison} shows the LeakyHammer covert channel capacity and error probability \nbcr{1}{(as defined in~\secref{sec:methodology})} for \nb{2 RFMs~(a) and 1 RFM~(b) per back-off across increasing noise intensities (~\secref{sec:prac-attack}). On the primary (left) y-axis, we plot error probability (blue line), and on the secondary (right) y-axis, we plot channel capacity (red line).} We make four key observations. 
First, \nbcr{1}{with 2-RFM back-offs (subfigure a), \X{} achieves 0.04 error probability and 29.95 Kbps channel capacity, at the lowest noise intensity.} \nbcr{1}{Compared to 4-RFM back-offs (\secref{sec:case_one}), error probability is higher across many noise intensities because the shorter latency of a 2-RFM back-off increases the likelihood that \X{} misdetects back-offs.  Second, with 1-RFM back-offs (subfigure b), \X{} exhibits higher error probability and lower channel capacity than both 2-RFM and 4-RFM back-offs across all noise intensities. This is because 1-RFM back-off latency partially overlaps with the periodic refresh latency, which prevents \X{} from reliably distinguishing back-offs. Third, as noise intensity increases, \X{}'s channel capacity decreases due to the increasing error probability with 2-RFM and 1-RFM back-offs. As the latency difference between a back-off and a periodic refresh decreases (e.g., from 4-RFM to 2-RFM to 1-RFM), \X{}'s channel capacity decreases due to the increasing error probability.}

{Fourth, to achieve higher channel capacity, an attacker can modify the covert channel attack to 1)~increase the transmission window to capture multiple potential back-offs, and 2)~filter periodic refreshes by calculating the time between each potential back-off and comparing it to the refresh interval. We evaluate this attack and observe that the channel capacity in the 1-RFM case is 21.53 Kbps at the lowest noise intensity (not shown in the figure).}
{We conclude that overlapping back-off and periodic refresh latencies degrade the channel capacity due to the increased error probability. However, the attacker can \nbcr{1}{increase} the channel capacity by modifying the attack.}

\subsection{{Effect of Preventive Action Latency}}
\label{sec:reduced_latency} 

\begin{figure}[h]
    \centering
    \includegraphics[width=0.9\linewidth]{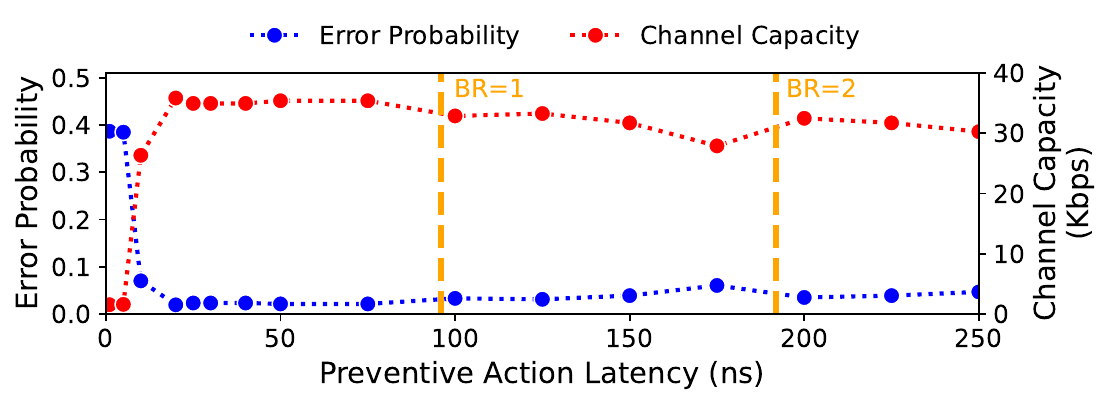}
    \caption{Error probability and channel capacity of LeakyHammer versus preventive action latency.}
    \label{fig:reduced_latency}
\end{figure}

We evaluate the channel capacity of \X{} with preventive action latencies smaller than \agy{the latency of an} \gls{rfm} \agy{command} \nb{in a system with a \gls{prac}-based defense configured with varying back-off latencies}.
\figref{fig:reduced_latency} shows the impact of preventive action latency on \X{}'s channel capacity and error probability as a scatter plot. The x-axis shows the preventive action latency (i.e., back-off latency) from \SI{0}{\nano\second} to \SI{250}{\nano\second}. The primary (left) and the secondary (right) y-axes show \X{}'s error probability and channel capacity and are marked by the blue and red data points, respectively.
\nbcr{1}{The two \agy{vertical} orange lines represent \emph{the minimum latency for a refresh-based preventive action}, showing the time needed to preventively refresh one aggressor's potential victim rows (i.e., 96~ns and 192~ns for a blast radius of 1 and 2~\cite{jedec2024jesd795c}).}

We make two observations \agy{from \figref{fig:reduced_latency}}.
First, reducing the preventive action latency to the minimum latency for preventive action does \agy{\emph{not}} \nbcr{1}{eliminate} the timing channel, \nbcr{1}{since} it \nbcr{1}{still} causes an observable latency difference.
Second, reducing the preventive action latency \nbcr{1}{eliminates} the timing channel \agy{\emph{only}} for \nbcr{1}{preventive action latencies} smaller than \agy{\SI{10}{\nano\second}}. This is because the preventive action always happens after an activation (i.e., only after a row buffer miss or a conflict), and any latency value greater than \SI{10}{\nano\second} helps \nbcr{1}{the attacker to} distinguish a preventive action from a row buffer conflict.  
We conclude that reducing
the preventive action latency \agy{\emph{only}} \nbcr{1}{eliminates} LeakyHammer for values smaller than \SI{10}{\nano\second}, which is far below the minimum latency required for a preventive action.

\subsection{\revcommon{Cache Hierarchy and Data Prefetching}} \labelcq{\label{resp:cq4}4\\D, E}
\label{sec:cache}
\revcommon{We evaluate LeakyHammer in a system with a larger cache hierarchy: 256 KB L2 and a 6 MB (per core) LLC, and enabling Best-Offset Prefetching~\cite{michaud2016best} at L2 cache. With the larger cache hierarchy, PRAC and RFM-based covert channels provide 36.7 (5.8\% reduction) and 47.7 Kbps (2.1\% reduction) channel capacity due to the increased cache access (and bypass) latency. The website fingerprinting attack provides 71.8\% (4.2\% lower) classification accuracy due to two reasons. First, the larger LLC filters some of the memory accesses and reduces the number of RowHammer preventive actions caused by the browser. Second, the data prefetcher incurs additional memory accesses that introduce noise. We conclude that a larger cache hierarchy and data prefetching do \emph{not} prevent LeakyHammer and \emph{only} have a \nbcr{1}{smaller} impact on channel capacity and accuracy. }

%% file: sections/06_discussion.tex
\section{Mitigating \X{}}
\label{sec:mitigations}

\atb{\agy{Any} secure \rh{} defense mechanism that performs
a preventive action based on an estimated
or precisely-tracked aggressor row activation
count (e.g.,~\mitigatingRowHammerCounterCitations{}) introduces a new timing 
channel \agy{that \X{} can exploit (\secref{sec:case_one},~\secref{sec:case_two},\secref{sec:sidechannel}).}\footnote{\atb{Probabilistic
\rh{} defense mechanisms (e.g., PARA~\cite{kim2014flipping}) and
mechanisms that hide the latency
of their preventive actions behind that of a periodic refresh
operation (e.g., TRR~\cite{micronddr4trr,frigo2020trrespass,cojocar2019eccploit,kim2020revisiting}) \nbcr{1}{might} \nbcr{1}{be challenging to exploit due to the attacker being unable to trigger and observe preventive actions reliably}. We discuss these defense mechanisms in detail
in~\secref{sec:other-mechanisms}.}}}
\nb{\agy{This section \nbcr{1}{introduces and evaluates}} \param{three} \X{} countermeasures\agy{: 1)\emph{~\gls{pmitig}},} a fundamental solution that decouples preventive actions from memory access patterns of running applications \agy{and 2)~\emph{Randomly Initialized Activation Counters (RIAC)},} a lower-cost countermeasure that reduces the channel capacity of \X{} at very low \rh{} thresholds (e.g., 64) where \nbcr{1}{\gls{pmitig}} incurs large performance overheads, and 3)~\emph{Bank-Level PRAC} that reduces the scope of the \X{} attacks to that of same-bank attacks (e.g., row buffer-based attacks~\cite{pessl2016drama,dramaqueen}) by performing per-bank preventive actions instead of rank or bank group-level preventive actions.}

\input{sections/06_01_periodic_victim_refresh}

\subsection{Introducing Noise to Reduce Channel\\Capacity}
\label{sec:riac}

Fundamentally mitigating \X{} \nbcr{1}{with \gls{pmitig}}
incurs high performance overheads at very low \rh{} thresholds (e.g., 128) \agy{(\secref{sec:mitig-eval}).}
\atb{An alternative solution is to reduce} \agy{\X{}'s channel capacity.}

\nbcr{1}{The key idea of randomly initialized activation counters (RIAC) is to introduce unintentional preventive actions at random activation counts to reduce the reliability of the covert channel, thereby decreasing channel capacity. At boot time RIAC initializes each activation counter with a random value instead of zero. Similarly, when an aggressor row triggers a preventive action, RIAC resets its activation counter value with a random value after the preventive action.} 
\nbcr{1}{By doing so, it introduces noise to the receiver's measurements by 1) preventing the sender and receiver from reliably determining when a preventive action will be triggered, and 2) causing unintentional preventive actions with fewer activations from concurrently running applications when counters are initialized close to the threshold.}
For example, for a \gls{prac}-based attack, if a counter is initialized close to the threshold, the receiver (or other applications in noisy environments) can trigger a back-off even when the sender is inactive (i.e., when transmitting 0). This effectively reduces the channel capacity of the attack (as we show in~\secref{sec:mitig-eval}).

We implement \atb{RIAC} for \gls{prac} (\pracrand{}) by initializing per-row activation counters randomly at boot time \textit{and} after each preventive action.\footnote{For this, \pracrand{} can generate random numbers in DRAM using a DRAM-based random number generator~\cite{kim2019d,olgun2021quac}. This can result in different random numbers for each chip for the same row, which we include in our experiments.}

\subsection{\reva{Bank-Level PRAC\labela{\label{resp:a1}{1}}}}
\label{sec:pracbank}
\reva{Another approach is to reduce the attack scope of \X{} to that of same-bank attacks (e.g., row buffer-based attacks~\cite{pessl2016drama,dramaqueen}) by making \rh{} preventive actions \nbcr{1}{finer-grained} (i.e., affecting only one bank instead of a channel or a bank group). \nbcr{1}{The key idea is to ensure preventive actions are visible only within the bank where they occur, and attackers accessing data outside the victim’s bank cannot observe them.}
\nbcr{1}{Supporting bank-level preventive actions} requires changes in the latest industry solutions (e.g., PRAC~\cite{jedec2024jesd795c} and RFM~\cite{jedec2020jesd795}). We use PRAC as an example to discuss such changes.}
A DRAM module needs to inform the memory controller about which bank requires a preventive action (e.g., with per-bank back-off signals), \nbcr{1}{allowing} the controller to issue memory requests to other DRAM banks during a back-off. \nbcr{1}{Bank-level preventive actions can be implemented in various ways,} such as increasing the number of back-off signals at the cost of increasing DRAM pins (i.e., ALERT\_n pins) or reading the bank information from the DRAM module, which requires changes in DRAM. 

\reva{We implement \nbcr{1}{a version of} bank-level PRAC (\textit{PRAC-Bank}) assuming each bank has an individual back-off signal.}
PRAC-Bank prevents attackers outside the victim’s bank from observing preventive actions, thereby reducing the attack scope of \X{} to same-bank covert channels and side channels. However, PRAC-Bank does not eliminate \X{}, since an attacker colocated in the same bank can still observe preventive actions.

\input{sections/06_02_mitig_eval}

%% file: sections/06_01_periodic_victim_refresh.tex
\subsection{\pmitiglong{} (FR-RFM)}
\label{sec:fr-rfm}

\gls{pmitig}'s key idea is to \nbcr{1}{decouple preventive actions from memory access patterns by} \atb{periodically} 
performing \atb{a preventive action}.
\agy{\gls{pmitig} issues an RFM command based on a \nbcr{1}{fixed} time period, unlike the 
periodic RFM (PRFM) mechanism (\secref{sec:case_two}) that issues an RFM command after a number of row activations targeting a DRAM bank as described in prior works \nbcr{1}{and the JEDEC standard}~\cite{jedec2020jesd795,canpolat2024understanding, canpolat2025chronus}.}
\nbcr{1}{With \gls{pmitig}}, \atb{a \agy{RowHammer-}preventive action \emph{cannot} leak any information about 
\nb{the row activations} of \agy{\emph{any}} concurrently running application. 
\nb{Thus, a sender cannot deterministically cause preventive actions, and an observer cannot observe if another application exhibits a memory access pattern that would lead to a preventive action.}
}

We \atb{design} \gls{pmitig} \atb{building} on the basics of \gls{prfm}.
\gls{prfm} \atb{securely} prevents \rh{} bitflips at \gls{nrh} by issuing an \gls{rfm} command once \atb{every} \gls{rfmth} row activations in a DRAM bank.\footnote{\atb{We refer the interested reader to~\cite{canpolat2024understanding,canpolat2025chronus,marazzi2022protrr} for \gls{prfm}'s security analysis.}} 
We set \gls{pmitigth} as the \agy{\emph{shortest time window needed}} to perform \gls{rfmth} row activations \agy{(i.e., \gls{pmitigth}=\gls{rfmth}$\times$\gls{trc}).}
\atb{Doing so ensures that \gls{pmitig} 1)~securely prevents \rh{} bitflips at \gls{nrh} because the memory controller \nbcr{1}{\textit{cannot}} issue more than \gls{rfmth} activate commands between two \gls{rfm} commands and 2)~prevents \X{} because the memory controller issues preventive actions (\gls{rfm} commands) precisely at the end of every fixed \gls{pmitigth} time interval independently from the memory access patterns of running applications.} \nb{To precisely schedule an \gls{rfm} command at \nbcr{1}{a} desired time \nbcr{1}{\emph{t}}, we modify the scheduler to ensure \nbcr{1}{that} all scheduled memory requests \nbcr{1}{complete and the active bank is precharged before \emph{t}.}}

\head{Security Analysis}
\agy{We define a sender and a receiver process that should \emph{not} be able to communicate with each other, but try to create a covert channel based on the number of RowHammer-preventive actions in a system that implements FR-RFM.}

\agy{Let $Req_{S}[i]$ and $Resp_{R}[i]$ represent the number of sender process's memory requests and the number of preventive actions observed by the receiver during the time window $i$, respectively.}
To ensure security, the number of preventive actions observed by the receiver within a window should \agy{\emph{not} change based on} the sender requests in that window. \agy{In other words $Resp_{R}[i]$ and $Req_{S}[i]$ are independent from each other.} 

\agy{By definition, FR-RFM issues an RFM command \nbcr{1}{at a fixed interval of} $T_{FRRFM}$, so the number of preventive actions that FR-RFM issues in a time window $i$ ($T_i$) is fixed at $T_{i}/T_{FRRFM}$. Therefore, the receiver \emph{cannot} observe more preventive actions than this fixed number.}\footnote{The receiver \emph{can} observe fewer \gls{rfm}s only when it fails to capture one or multiple \gls{rfm}s within the window. This can happen due to contention in memory, which is within the scope of memory contention-based attacks~\memorycontentionattacks{} and outside of the scope of this paper.}

%% file: sections/06_02_mitig_eval.tex
\subsection{Evaluation of \X{}\\Countermeasures}
\label{sec:mitig-eval}

\head{Channel Capacity Reduction}
\agy{We evaluate \X{}'s channel capacity in systems that employ \nb{\gls{pmitig} and \pracrand{}}\footnote{PRAC-Bank does not affect the PRAC-based \X{} covert channel described in our work.} following the methodology described in \secref{sec:methodology} where \nb{the sender transmits messages with different patterns \nbcr{1}{(as explained in~\secref{sec:case_one})}}. Our evaluation shows that \gls{pmitig} and \pracrand{} reduce \X{}'s channel capacity by \SI{100}{\percent} and \SI{86}{\percent}, respectively, on average across \nb{all tested message patterns}.
\gls{pmitig} completely \nbcr{1}{eliminates the channel} by decoupling preventive actions from memory access patterns. 
\pracrand{} introduces significant noise to \X{} by randomizing activation counters and causing them to reach the threshold of triggering \nb{back-offs} sooner than \gls{prac}'s counters.\footnote{\nbcr{1}{The magnitude of reduction in channel capacity depends on (i)~the attack implementation, (ii)~random initialization values of the activation counters, and (iii)~memory access patterns.}}}

\begin{figure}[!h]
    \centering
    \includegraphics[width=0.95\linewidth]{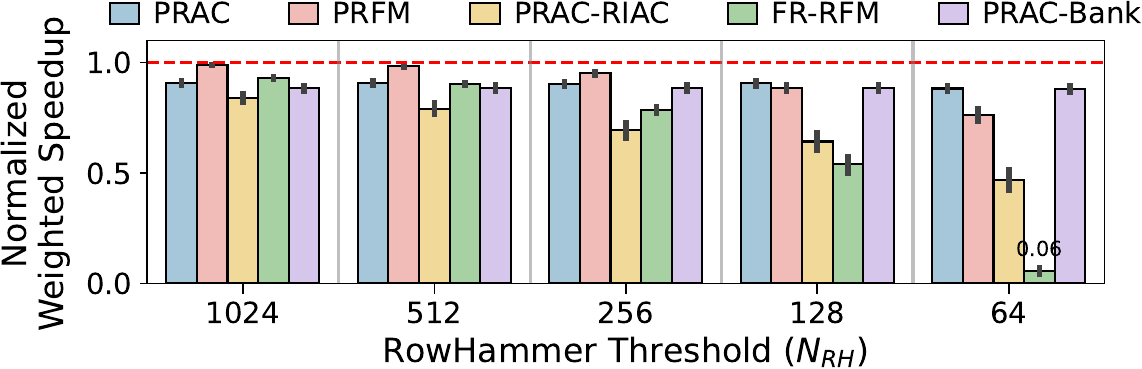}
    \caption{Performance of PRAC, RFM and \X{} countermeasures \nbcr{1}{(\pracrand{}, \gls{pmitig}, and PRAC-Bank)} normalized
    to a baseline system with \emph{no} \rh{} or \X{} mitigation.}
    \label{fig:mitig-overhead}
\end{figure}

\head{Performance Overhead}
\figref{fig:mitig-overhead}
shows performance (in terms of
weighted speedup) for PRAC (\secref{sec:case_one}), PRFM (\secref{sec:case_two}), \pracrand{} (\secref{sec:riac}), \gls{pmitig} (\secref{sec:fr-rfm}), \reva{and PRAC-Bank (\secref{sec:pracbank})} over \param{five} RowHammer threshold values (x-axis)
normalized to a baseline system that
does \emph{not} implement \rh{} and \X{} mitigations \nb{for \param{60} four-core workloads consisting of SPEC2017~\cite{spec2017} and SPEC2006~\cite{spec2006} applications.}

Based on these results, we make \param{six} observations.
First, \gls{pmitig} fundamentally solves \X{} while incurring \param{7.0\%} average performance overhead at \gls{nrh}$=1024$, compared to a system with \agy{\emph{no}} \rh{} mitigation.
Second, \gls{pmitig} performs \nbcr{1}{similarly} to \gls{prac} and \gls{prfm} at \gls{nrh}$\ge512$, while slightly outperforming \gls{prac} at \gls{nrh}$=1024$. 
Third, \pracrand{} incurs \param{16.0\%} and \param{35.8\%} average performance overheads at \gls{nrh}$=1024$ and \gls{nrh}$=128$, compared to no \rh{} mitigation.
Fourth, \pracrand{} outperforms \gls{pmitig} starting from \gls{nrh}$=128$. This is \nbcr{1}{because} at \gls{nrh}$<256$, \gls{rfm}-based mitigations incur higher \agy{performance} overheads due to performing preventive refreshes \agy{more \nbcr{1}{frequently} (i.e.,} with very low \gls{rfmth} values\agy{)} to ensure security against \rh{}. 
Fifth, \agy{under the extreme condition of \nbcr{1}{\gls{nrh}=}64, \gls{pmitig}'s and \pracrand{}'s performance overheads reach \revb{$18.2\times$} and \revb{$2.14\times$}, \labelb{\label{resp:b2}2}
respectively, on average, across all tested workloads.}
\reva{Sixth, PRAC-Bank performs similarly to PRAC across all \gls{nrh} values (within 2.5\%), and we do not observe any significant change in performance due to making back-offs finer-grained.}
\agy{From these observations, we} conclude that \gls{pmitig} \nbcr{1}{completely} prevents \X{} while performing similarly to \rh{} mitigations that are insecure against \X{}. However, completely eliminating the timing channel with \gls{pmitig} incurs \nb{large} performance overheads at very low \rh{} thresholds \nbcr{1}{(i.e., \gls{nrh}$\le$128)} due to \agy{performing preventive refreshes more \nbcr{1}{frequently} than needed to prevent RowHammer bitflips. Hence, for very low \rh{} thresholds, defending against \rh{} defense-based timing channels remains an open and critical research problem. We hope and expect that our \nbcr{1}{analyses and} findings will inspire and help future research to address this problem.}

%% file: sections/07_othermechs.tex
\section{\X{} with Other \rh{}\\Mitigations}
\label{sec:other-mechanisms}
\ous{A RowHammer defense-based timing channel is possible when an attacker can both i)~observe a preventive action\nb{'s overhead} and ii)~\nbcr{1}{intentionally} trigger a preventive action.
Thus, the channel depends on two properties of the mechanism: the preventive action and the trigger algorithm.
We classify two types of preventive actions: 1) \emph{\nb{Overlapped} Latency} and 2) \emph{Observable Latency}.}

\head{\nb{Overlapped} Latency}
\ous{Several works~\cite{qureshi2024mint, jaleel2024pride, hassan2021utrr, hong2023dsac} use \emph{borrowed time} from periodic refreshes~\cite{hassan2021utrr,chang2014improving} to \nb{perform preventive actions}.
These mechanisms \nb{hide} preventive action latency.
However, performing preventive actions \textit{only} \nbcr{1}{during} periodic refreshes limits \nb{the rate of preventive refreshes that can be performed and ensures robustness only for} \gls{nrh} \nbcr{1}{values} of several thousands~\cite{qureshi2024mint, jaleel2024pride, hong2023dsac}, which is much higher than the values \atb{evaluated} by recent works~\cite{olgun2024abacus, canpolat2024understanding, bostanci2024comet, canpolat2025chronus}.}

\head{Observable Latency}
\ous{Observable latency preventive actions include 1)~preventively refreshing a victim row~\refreshBasedRowHammerDefenseCitations{}, 2)~dynamically remapping an aggressor row~\cite{saileshwar2022randomized, wi2023shadow, woo2023scalable, saxena2022aqua}, and 3)~throttling unsafe accesses~\cite{yaglikci2021blockhammer, blockhammergithub,canpolat2024breakhammer}.
Mechanisms using these actions execute a \emph{trigger algorithm} that determines when to perform the preventive action.
We discuss three classes of trigger algorithms: \emph{exact}, \emph{approximate}, and \emph{random}.
First, exact trigger algorithms guarantee perfect tracking (\nbcr{1}{of} activations performed on a row or bank) by allocating a \nbcr{1}{tracker} for each resource (e.g., row or bank) and perform actions based on these trackers~\cite{jedec2024jesd795c, yaglikci2021security}.
These mechanisms enable an attacker to reliably trigger and observe preventive actions (\secref{sec:case_one},~\secref{sec:case_two},~\secref{sec:sidechannel}).
Second, approximate trigger algorithms allocate \atb{fewer} \nbcr{1}{trackers} than \nbcr{1}{exact tracking algorithms} and use frequent item counting \atb{algorithms} \nbcr{1}{or grouping methods} to approximate \nbcr{1}{the activation counts}.
Mechanisms using approximate trigger algorithms~\cite{park2020graphene, qureshi2022hydra, saileshwar2022randomized, saxena2022aqua} can potentially increase the noise in a RowHammer defense-based timing channel.
This is because the memory accesses of different processes can share \nbcr{1}{trackers} used by the attack, unpredictably \nbcr{1}{influencing} when a preventive action is triggered.
Third, random trigger algorithms are \emph{stateless}, and they perform preventive actions based on a random number.
These mechanisms (e.g., PARA~\cite{kim2014flipping}) make it challenging to create RowHammer \nbcr{1}{defense}-based timing channels because an attacker \emph{cannot} reliably trigger or observe preventive actions, at the cost of increased performance overhead~\cite{yaglikci2022hira, kim2020revisiting,olgun2024abacus,bostanci2024comet}.}

%% file: sections/08_relatedwork.tex
\section{Related Work}
This paper \nbcr{1}{is the first to introduce a new class of} timing covert channel and side channel vulnerabilities \nbcr{2}{(called \X{}) that stem from} \rh{} {defenses}. \nbcr{0}{We already discuss and compare \X{} against the closely related works~\cite{dramaqueen,dramaqueen,maurice2017hello,yarom2014flush+,gruss2016flush+,liu2015last,xiong2020leaking,lipp2020take} in~\secref{sec:comparison_against_others}. In this section, we discuss other relevant prior \nbcr{1}{research}.}

\head{Main Memory-based and Cache-Based \nbcr{1}{Covert Channel and Side Channel} Attacks}
Prior works propose timing \nbcr{1}{covert channel and side channel} attacks exploiting main memory and cache-based structures. % (e.g., cache sets, row buffer, and scheduler queues). 
A set of prior works exploits the DRAM row buffer to leak private information and build covert communication channels \nbcr{1}{(as discussed in detail in~\secref{sec:comparison_against_others})}. 
DRAMA~\cite{pessl2016drama}, DRAMAQueen~\cite{dramaqueen}, and IMPACT~\cite{bostanci2025revisiting} exploit the DRAM row buffer states for building timing attacks. \nbcr{1}{Various} prior works (e.g., ~\cite{pessl2016drama,xiao2016one, kwong2020rambleed,moscibroda2007memory,bhattacharya2016curious}) leak DRAM address mappings exploiting the row buffer-based timing differences.
Other \agy{main memory-based} attacks exploit the memory contention~\memorycontentionattacks{}, memory deduplication~\cite{xiao2013security,lindemann2018memory} and (de)compression~\cite{kelsey2002compression,rizzo2012crime,gluck2013breach,be2013perfect,vanhoef2016heist,van2016request,karakostas2016practical,tsai2020safecracker,schwarzl2021practical} latencies. 
Several works exploit the cache timing variation~\cite{maurice2017hello,yarom2014flush+,gruss2016flush+,liu2015last,xiong2020leaking,lipp2020take}. \X{} is a new attack class that exploits \rh{} defenses' preventive actions to create timing channels. 

{The idea of exploiting industry solutions to \rh{} was developed independently and concurrently\footnote{An earlier version of our paper was submitted to ISCA 2025~\cite{bostanci2025understanding}.} by \nbcr{2}{three} recent works~\cite{woo2025mitigations,nazaraliyev2025not,taneja2025roguerfm}. These \nbcr{2}{three} works evaluate attacks on one \rh{} defense \nbcr{1}{(i.e., PRAC~\cite{jedec2024jesd795c,canpolat2024understanding} in~\cite{woo2025mitigations}, and RFM~\cite{jedec2020jesd795} in~\cite{nazaraliyev2025not,taneja2025roguerfm})}. In contrast, our work proposes and evaluates attacks exploiting two of the latest industry solutions to \rh{} and analyzes prior \rh{} defenses from academia and industry in terms of timing channel leakage. We believe our work and these concurrent works highlight the \nbcr{1}{prominence and} importance of timing channels introduced by \rh{} defenses and the need for low-cost mitigations against \X{}, as demonstrated by our analyses and results.}

\head{Main Memory-based Timing Channel Defenses}
We already \nbcr{1}{describe} and evaluate \nbcr{1}{three} countermeasures that fundamentally prevent \X{} timing channel, reduce its channel capacity, and reduce the attack scope to that of existing main memory-based attacks (\secref{sec:mitigations}).
Another approach is to restrict the fine-grained timers' usage to prevent attackers from measuring memory requests' latencies, which is applied in some modern processors \atb{(e.g., Apple M1~\cite{ravichandran2022pacman})}. However, these timers are used in many userspace applications, and restricting their usage can disable \nbcr{1}{these applications} or degrade their \atb{performance}.
DAGguise~\cite{deutsch2022dagguise} \atb{and Camouflage~\cite{zhou2017camouflage} defend} 
against memory timing 
side-channel attacks \atb{by} obfuscat\atb{ing} an application’s memory access patterns.
\atb{Although these two techniques do \emph{not} mitigate the \rh{} defense-based timing channels, they can be adapted
to mitigate side channel attacks that exploit \rh{} preventive actions.}

%% file: sections/09_conclusion.tex
\section{Conclusion}

{We present the first analysis and evaluation of timing channel vulnerabilities introduced by \rh{} defenses.}
\agy{Our key observation is that} \rh{} defenses' \emph{preventive actions} 
\agy{have two fundamental features {that} \atb{allow an attacker to exploit
\rh{} defenses for timing channels:}} \ous{1)~preventive actions often reduce DRAM bandwidth availability, resulting in large \nbcr{1}{and easily measurable} memory access latencies}, 2)~a user can intentionally trigger a {\agy{preventive action} because preventive actions highly depend on memory access patterns.}
{We introduce} \X{}, a new class of attacks that leverage \rh{} defense-induced memory latency differences. We demonstrate \X{} by building covert channel and side channel attacks. 
First, we build \atb{\param{two}} covert communication channels exploiting \atb{\param{two}} {state-of-the-art \rh{} defenses \nbcr{1}{adopted widely in industry}.}
Second, we demonstrate a proof-of-concept website fingerprinting attack
{that can identify {visited websites}.} We propose \param{three} \X{} countermeasures.
\nbcr{1}{One countermeasure fully eliminates the \X{} timing channel with low overhead at near-future \rh{} thresholds, but incurs high costs at very low \rh{} thresholds. In such cases, countermeasures that reduce channel capacity are more practical due to their lower overhead. We therefore call for further research on \nbcr{2}{better solutions and more robust systems in the presence of such vulnerabilities.}}

%% file: sections/20_artifact.tex
\section{Artifact Appendix}
\nobalance
\subsection{Abstract}

This artifact provides the source code and scripts to reproduce the key experiments and their results presented in our MICRO 2025 paper. This artifact enables reproducing the following key figures:

\begin{enumerate}
\item PRAC-induced high memory access latencies and their observability (Figure 2)
\item PRAC-based covert channel attack demonstrating 40-bit message transmission (Figure 3)
\item PRAC-based covert channel's capacity and error probability with increasing noise intensities (Figure 4)
\item RFM-based covert channel attack demonstrating 40-bit message transmission (Figure 6)
\item RFM-based covert channel's capacity and error probability with increasing noise intensities (Figure 7)
\end{enumerate}

If artifact evaluation time permits, we will extend the artifact with more results such as LeakyHammer mitigation evaluation.
\subsection{Artifact Checklist (Meta-information)}

{\small
\begin{itemize}
    \item {\bf Program:} C++ programs, Python scripts, shell scripts.
    \item {\bf Compilation:} cmake, GNU make, scons.
    \item {\bf Run-time environment: } Linux (tested on Ubuntu 20.04 and 22.04 with the provided container), Python 3.8
    \item {\bf Execution: } Bash scripts, Python scripts.
    \item {\bf Metrics: } Timing results, performance metrics of attacks and defenses.
    \item {\bf Output: } Figures in PDF format and related data in plaintext files.
    \item {\bf How much disk space required (approximately)?: } 10GB.
    \item {\bf How much time is needed to prepare workflow (approximately)?: } $\sim1$ hour.
    \item {\bf How much time is needed to complete experiments (approximately)?:} 2-4  hours.
    \item {\bf Publicly available?: } Yes
    \item {\bf Archived (provide DOI)?: } Yes, DOI: https://doi.org/10.5281/zenodo.16734696
\end{itemize}
}

\subsection{Description}
\subsubsection{How to Access} The source code and scripts can be downloaded from Zenodo (\url{https://doi.org/10.5281/zenodo.16734696}).

\subsubsection{Hardware dependencies}

If the reader will be running the experiments in their own systems or compute clusters\footnote{To enable easy reproduction of our results, we can provide SSH access to our internal Slurm-based infrastructure with all the required hardware and software during the artifact evaluation. Please contact us through HotCRP and/or the AE committee for details.}:
\begin{itemize}
\item We will be using Docker images to execute experiments. These Docker images assume x86-64 systems.
\item The experiments have been executed using a Slurm-based infrastructure. We \textbf{strongly} suggest using such an infrastructure for bulk experimentation due to the number of experiments required. However, our artifact provides scripts for 1) Slurm-based and 2) local execution.
\item Each experiment takes around 1 hour, depending on the experiment. 
\item At least 16GB of RAM (higher is better for parallel experiments).
\item \nbcr{1}{Having a large set of compute nodes (e.g., 64 cores) for parallel experiments reduces the total time to reproduce the results, and hence, is strongly suggested.}
\end{itemize}

\subsubsection{Software dependencies}
\begin{itemize}
    \item Linux Operating System (tested on Ubuntu 20.04 and 22.04 with the provided container).
    \item Docker or Podman
    \item Bash.
    \item Python3 (version 3.8)
    \item Packages: pandas numpy matplotlib seaborn pyyaml scipy scons
    \item CMake, GNU Make 
    \item wget, tar.
    \item GCC (10.5.0), g++ (10.5.0). 
    \item zlib1g zlib1g-dev libprotobuf-dev protobuf-compiler libprotoc-dev libgoogle-perftools-dev build-essential m4
    \item Slurm (strong recommendation)

\end{itemize}

We \textbf{strongly recommend} using our container image to satisfy all dependencies with the correct versions selected. Our scripts already include instructions to build the image from our Dockerfile. To build it manually with \texttt{docker} or \texttt{podman}, run:

\fancycommand{\$ [docker/podman] build . $--$no-cache $--$pull -t leakyhammer\_artifact}

To install all requirements manually (to run in a native environment\footnote{We provide ae\_install\_requirements.sh script to help get started in a native environment. For artifact evaluation purposes (i.e., reproduce all figures exactly as they are presented in the paper), we strongly recommend using the container-based execution instructions. These instructions are already given as default in this appendix.}):
\fancycommand{\$ ./ae\_install\_requirements.sh}

To install only python3 dependencies with pip manually:

\fancycommand{python3 -m pip install -r requirements.txt}

\subsection{Installation}

The following steps downloads and prepares the repository for the main experiments. We assume \texttt{podman} is the selected container tool. It is also possible to use \texttt{docker} by typing \texttt{docker} where \texttt{podman} is given. \footnote{To run all experiments without the container image for the artifact please check our README.md for instructions.}

\textbf{1. Getting started:}
\fancycommand{\$ cd LeakyHammer}

\textbf{2. Set up the container, build simulators and run quick experiments:}

\fancycommand{\$ ./container\_setup.sh podman}

This script saves the container image as \texttt{"leakyhammer\_artifact.tar"} to use in future experiments.

\subsection{Experiment Workflow}

Our artifact contains 1) our source code of gem5 and Ramulator2, 2) LeakyHammer timing attack, latency measurement, and noise generator scripts, 3) experiment automization scripts for local and Slurm-based infrastructures, and 3) python scripts to plot all figures and print results.

The following steps runs gem5 experiments in parallel (in local environment or using Slurm):

\textbf{1. Review and update configurations based on environment (e.g., the number of parallel jobs, Slurm username and partition name)}

To review (and update) script configurations set in the \texttt{gem5/result-scripts/run\_config.py} based on the selected execution environment: 

\fancycommand{\$ head -n36 ./gem5/result-scripts/run\_config.py\\vim ./gem5/result-scripts/run\_config.py}

This script enables configuring the number of concurrently running jobs for slurm and local execution environments, user and partition names for slurm-based execution, and various options to modify the experiments (e.g., the number of bytes transmitted for channel capacity experiments). 

\textbf{2. Run experiments:}

\textit{Using Slurm:}
\texttt{run\_parallel\_slurm\_container.sh} creates the \textbf{results} directory and \textbf{prac} and \textbf{rfm} subdirectories for the covert channel attacks. Within each subdirectory, there should be two directories \textbf{baseline}, and \textbf{noise} for different experiment results.

\fancycommand{\$ sh gem5/run\_parallel\_slurm\_container.sh podman}

The script submits Slurm jobs executing scripts created in \textbf{run\_scripts/}. Based on the maximum concurrently running (or scheduled) Slurm job limitation, it may stop and retry after an interval (configurable in \texttt{run\_config.py}). It terminates after submitting all jobs.

\textit{For local environment:}

Use the following command to run all experiments using python's ThreadPoolExecutor. The number of concurrent workers is configurable in \texttt{run\_config.py}.

\fancycommand{\$ sh gem5/run\_parallel\_local\_container.sh podman}

The script will run experiments in parallel using the resources defined in the \texttt{./gem5/result-scripts/run\_config.py} script and terminate when all experiments are completed. Using \texttt{tmux} or similar tools is recommended due to the high execution time.

\subsection{Plotting the Results}

To plot all figures using the container image:

\fancycommand{\$ sh gem5/plot\_figures\_container.sh podman}

These scripts parse all results and plot the figures. All generated figures are stored in the \texttt{gem5/figures} directory. The parsed results are saved in \texttt{gem5/results/} as csv files. After all experiments are completed, the plotting script is expected to create \texttt{ber\_[prac/rfm].csv} and \texttt{noise\_ber\_[prac/rfm].csv} files.
Finally, the scripts print out a summary of the results to the terminal. 

\subsection{Evaluation \& Expected Results}

Please review Figures \ref{fig:latency_prac}, \ref{fig:prac-poc}, \ref{fig:prac-channel-cap}, \ref{fig:rfm-poc} and \ref{fig:rfm-channel-cap} in the \texttt{gem5/figures/} directories.
Note that running the experiments in without the provided container image might compile the attack scripts with different compiler versions. This might result in slightly different data points shown in the figures based on your system configurations (e.g., compiler version) but it does not change the key observations.

The following results are printed at the end of the plotting script in the previous step.
\begin{enumerate}
    \item (Section \ref{claim:1}) We observe \agy{that} the PRAC-based attack achieves \textbf{\param{39.0 Kbps}} raw bit rate.
    \item (Section \ref{claim:2}) We observe that the RFM-based attack achieves \textbf{\param{48.7 Kbps}} raw bit rate. 
    \item (Section \ref{claim:3}) For PRAC-based attack: At the noise intensity 1\% (shown with the orange line), the channel capacity is \textbf{\param{28.8 Kbps}}.  
    \item (Section \ref{claim:3}) The PRAC-based covert channel's capacity remains high (>\textbf{20.7Kbps}) until a very high noise intensity value of 88\%. 
    \item (Section \ref{claim:4}) For RFM-based attack: At the noise intensity 1\% (shown with the orange line), the channel capacity is \textbf{\param{46.3 Kbps}}. 
    \item (Section \ref{claim:4}) The RFM-based covert channel's capacity remains high (>\textbf{20.7 Kbps}) until a noise intensity of 50\% (shown with the purple line).  
\end{enumerate}

\subsection{Troubleshooting}

To retry failing jobs (for any reason), the run script (e.g., \texttt{gem5/run\_parallel\_slurm\_container.sh}) can be run again with \texttt{SKIP\_EXISTING= True} option in the \texttt{run\_config.py} script.

\fancycommand{\$ sh gem5/run\_parallel\_local\_container.sh podman}

This way, the run script checks if a valid result is present in the output file for each experiment, and only resubmit/rerun an experiment if it failed previously. Note that, the run script should be executed \textit{after} all remaining jobs are completed to avoid disrupting existing jobs.

\subsection{Methodology}

Submission, reviewing and badging methodology:

\begin{itemize}
    \item \url{https://www.acm.org/publications/policies/artifact-review-and-badging-current}
    \item \url{http://cTuning.org/ae/submission-20201122.html}
    \item \url{http://cTuning.org/ae/reviewing-20201122.html}
\end{itemize}